\documentclass[preprint,12pt]{elsarticle}
\usepackage{amsmath}
\usepackage[flushleft]{threeparttable}
\usepackage{lineno,hyperref}
\usepackage{adjustbox}
\usepackage{nomencl}
\usepackage{etoolbox}
\usepackage{framed} 
\modulolinenumbers[5]
\journal{Journal of \LaTeX\ Templates}
\begin{document}
\begin{frontmatter}
\title{Heat transfer analysis in an uncoiled model of the cochlea during magnetic cochlear implant surgery}

\author{Fateme Esmailie}\author{Mathieu Francoeur}\author{Tim Ameel\corref{mycorrespondingauthor}}
\address{Department of Mechanical Engineering, University of Utah, Salt Lake City, Utah 84112, USA}
\cortext[mycorrespondingauthor]{Corresponding author}
\ead{ameel@mech.utah.edu}

\begin{abstract}
Magnetic cochlear implant surgery requires removal of a magnet via a heating process after implant insertion, which may cause thermal trauma within the ear. Intra-cochlear heat transfer analysis is required to ensure that the magnet removal phase is thermally safe. The objective of this work is to determine the safe range of input power density to detach the magnet without causing thermal trauma in the ear, and to analyze the effectiveness of natural convection with respect to conduction for removing the excess heat. A finite element model of an uncoiled cochlea, which is verified and validated, is applied to determine the range of maximum safe input power density to detach a 1-mm-long, 0.5-mm-diameter cylindrical magnet from the cochlear implant electrode array tip. It is shown that heat dissipation in the cochlea is primarily mediated by conduction through the electrode array. The electrode array simultaneously reduces natural convection due to the no-slip boundary condition on its surface and increases axial conduction in the cochlea. It is concluded that natural convection heat transfer in a cochlea during robotic cochlear implant surgery can be neglected. It is found that thermal trauma is avoided by applying a power density from $2.265\times10^7$ $\textup W/ \textup m^3$ for 114 s to $6.6\times10^7$ $\textup W/ \textup m^3$ for 9 s resulting in a maximum temperature increase of 6$^{\circ}$C on the magnet boundary.
\end{abstract}

\end{frontmatter}

\makenomenclature
\renewcommand\nomgroup[1]{%
  \item[\bfseries
  \ifstrequal{#1}{A}{}{%
  \ifstrequal{#1}{S}{Subscripts}{%
  \ifstrequal{#1}{G}{Greek Symbols}{}}}%
]}

\mbox{}

\nomenclature[01]{$b$}{Distance between two concentric cylinders [m]}
\nomenclature[02]{$c_p$}{Heat capacity [J/kg$\cdot$K]}
\nomenclature[03]{$D$}{Diameter [m]}
\nomenclature[04]{$\bf{g}$}{Gravitational acceleration [m/s$^{2}$]}
\nomenclature[05]{$\bf{I}$}{$3\times3$ identity matrix}
\nomenclature[06]{$k$}{Thermal conductivity [W/m$\cdot$K]}
\nomenclature[07]{$L$}{Gap distance between two circles [m]}
\nomenclature[08]{$Pe$}{P\'{e}clet number}
\nomenclature[09]{$Pr$}{Prandtl number}
\nomenclature[10]{$q$}{Input power density [W/m$^3$]}
\nomenclature[11]{$r$}{Radial direction}
\nomenclature[12]{$r'$}{Radial distance from the inner cylinder center [m]}
\nomenclature[13]{$R$}{Radius [m]}
\nomenclature[14]{$R'$}{Radial distance between the two cylinder walls [m]}
\nomenclature[15]{$R^*$}{Dimensionless radial distance $\frac{r-R_i}{R_o-R_i}$}
\nomenclature[16]{$R^{*'}$}{Dimensionless radial distance $\frac{r'-R_i}{R'-R_i}$}
\nomenclature[17]{$Ra$}{Raleigh number}
\nomenclature[18]{$T$}{Temperature [$^{\circ}$C]}
\nomenclature[19]{$t$}{Time [s]}
\nomenclature[20]{$\mathbf{u}$}{Velocity vector [m/s]}
\nomenclature[21]{$x$}{Direction along the electrode array}
\nomenclature[22]{$y$}{Horizontal direction orthogonal to the electrode array}
\nomenclature[23]{$z$}{Vertical directional orthogonal to the electrode array}
\nomenclature[S]{$bc$}{Body core}
\nomenclature[S]{$cc$}{Concentric cylinders}
\nomenclature[S]{$eff$}{Effective}
\nomenclature[S]{$i$}{Inner}
\nomenclature[S]{$o$}{Outer}
\nomenclature[S]{$1$}{Region 1 (Magnet)}
\nomenclature[S]{$2$}{Region 2 (Perilymph)}
\nomenclature[G]{$\alpha$}{Thermal diffusivity [m$^2$/s]}
\nomenclature[G]{$\varepsilon$}{Ratio of inner cylinder eccentricity to the gap distance between two cylinders in a concentric arrangement}
\nomenclature[G]{$\theta$}{Dimensionless temperature}
\nomenclature[G]{$\rho$}{Density [kg/m$^3$]}
\nomenclature[G]{$\varphi$}{Azimuthal angle [$^{\circ}$]}

\printnomenclature
    

\section{Introduction}
\begin{figure}[t]
\centering\includegraphics[width=\linewidth]{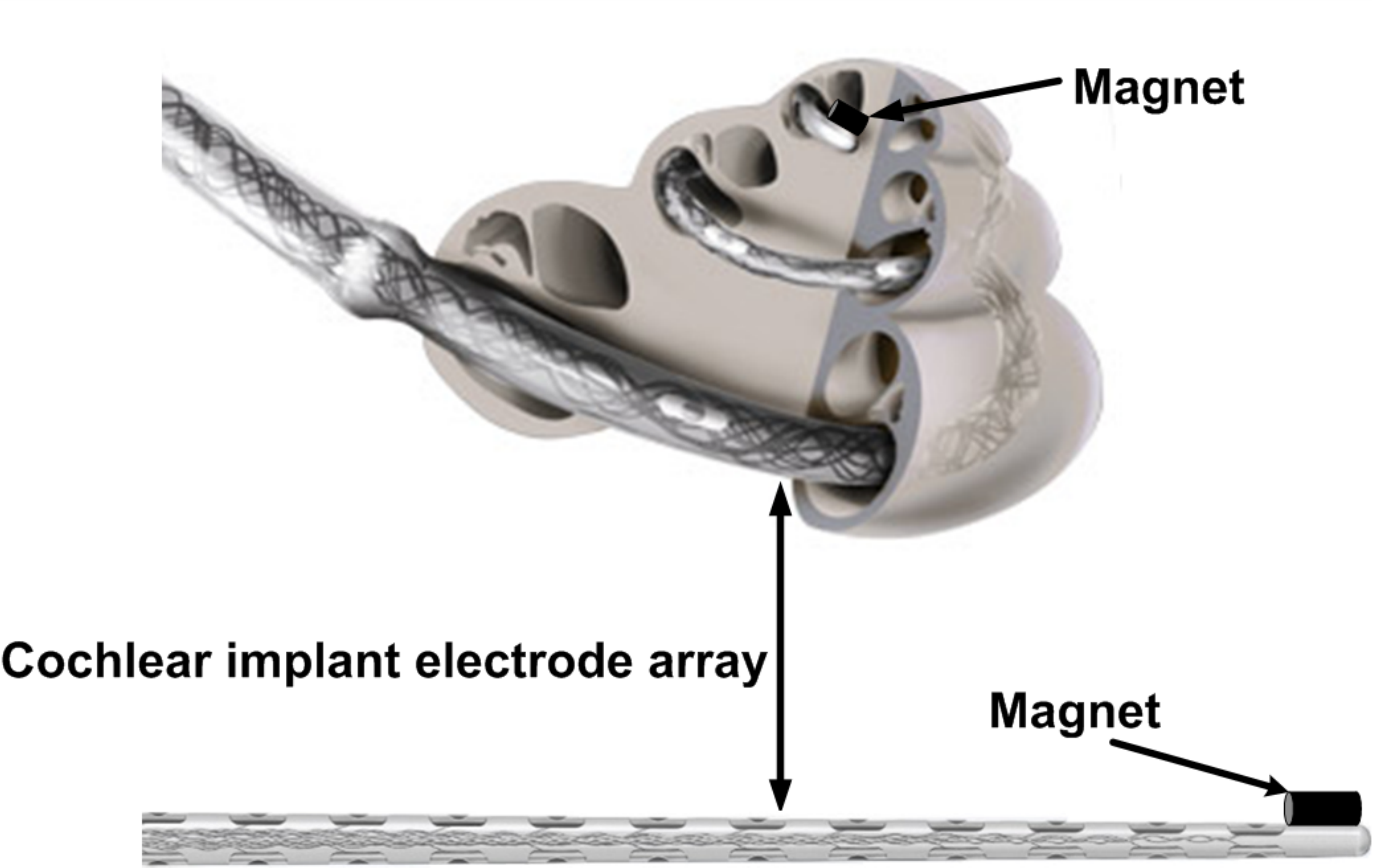}
\caption{\label{fig:1} Cutaway view of a cochlea with an inserted electrode array (Photo by MED-EL) \cite{DHANASINGH201793}, \cite{medel}.}
\end{figure}

A cochlear implant is a Food and Drug Administration (FDA) approved solution for profound-to-severe hearing disability. Manual insertion of cochlear implant electrode arrays (hereafter called electrode array), however, causes intra-cochlear physical trauma in about one-third of surgeries \cite{Clark}, \cite{Clark2}. This physical trauma not only decreases the residual hearing ability but also reduces the functionality of the cochlear implant \cite{Clark}, \cite{Clark2}. To prevent physical trauma during surgery, researchers have suggested magnetic guidance of the electrode array \cite{Clark}, \cite{Clark2}, \cite{Leon}. In this technique, a magnet attached to the tip of the electrode array is guided in the cochlear turns via an external magnetic field (see Fig.~\ref{fig:1}) \cite{DHANASINGH201793}, \cite{medel}. After surgery, the magnet must be detached from the electrode array and removed from the cochlea to avoid potential medical complications arising when the patient is exposed to a strong magnetic field \cite{majdani}. The detachment process requires heating of the magnet, thus releasing thermal energy in the cochlea that may cause thermal trauma within the ear. Heat transfer in the middle ear, organ of balance, auditory nerves and the skull has been studied for applications such as caloric test \cite{Kassemi:2005}, \cite{Baertschi1975}, \cite{Cawthorne}, stapedectomy \cite{morshed}, \cite{ricci_mazzoni_1985}, \cite{Kodali}, radio-frequency radiation \cite{McIntosh}, \cite{Bernardi}, \cite{McIntosh2005}, magnetic resonance imaging \cite{majdani}, \cite{Wanger}, \cite{LOEFFLER2007583}, \cite{YUN2005275}, infrared neural stimulation of cochlear implants \cite{THOMPSON201546}, \cite{Shiparo}, \cite{Izzo2007OpticalPV}, \cite{ Thompson2013InfraredNS} \cite{rajguru4-INS}, , and therapeutic hypothermia \cite{Rajguru}\cite{rajguru2-terapeutic}, \cite{rajguru3-therapeutic}. Yet, despite these numerous efforts, heat transfer analysis within cochlear canals is still lacking and constitutes an important knowledge gap to the establishment of magnetic cochlear implant surgery. In addition, neither the heat source nor the targeted tissue in these aforementioned applications are similar to the magnet detachment process after magnetic insertion of the cochlear implant. Clearly, a comprehensive thermal analysis in cochlear canals during magnetic cochlear implant surgery is required. This constitutes the novelty of this work, as heat transfer in cochlear canals has never been studied before.  \\ 

The objective of this paper is to understand the mechanisms responsible for thermal energy dissipation during magnetic guidance of cochlear implants. For that purpose, conduction and natural convection heat transfer are simulated in a three-dimensional (3D) uncoiled model of the cochlea, where the magnet acts as a heat source that results from Joule heating. Specifically, the safe range of input power density to detach the magnet without causing thermal trauma in the ear, and the effectiveness of natural convection with respect to conduction for removing the excess heat during the magnet detachment phase, are analyzed for the first time.\\

The rest of the paper is organized as follows. Section 2 provides a description of the computational model and the associated assumptions. Next, the model is verified for conduction heat transfer by comparison against a one-dimensional (1D) solution for two concentric cylinders, where the inner cylinder represents the magnet generating heat. This is followed by a verification of the model for natural convection heat transfer between two concentric cylinders and validation for two eccentric cylinders. In the fourth section, heat transfer within the uncoiled model of the cochlea where the magnet acts as a heat source is simulated. The impact of natural convection with and without an inserted electrode array are analyzed. The effect of the magnitude and duration of the heating process, and the fluid within the cochlear canals, are determined in the last section. Concluding remarks are then provided. \\

\section{Description of the 3D uncoiled model of the cochlea}
\begin{figure}[h]
\centering\includegraphics[width=\linewidth]{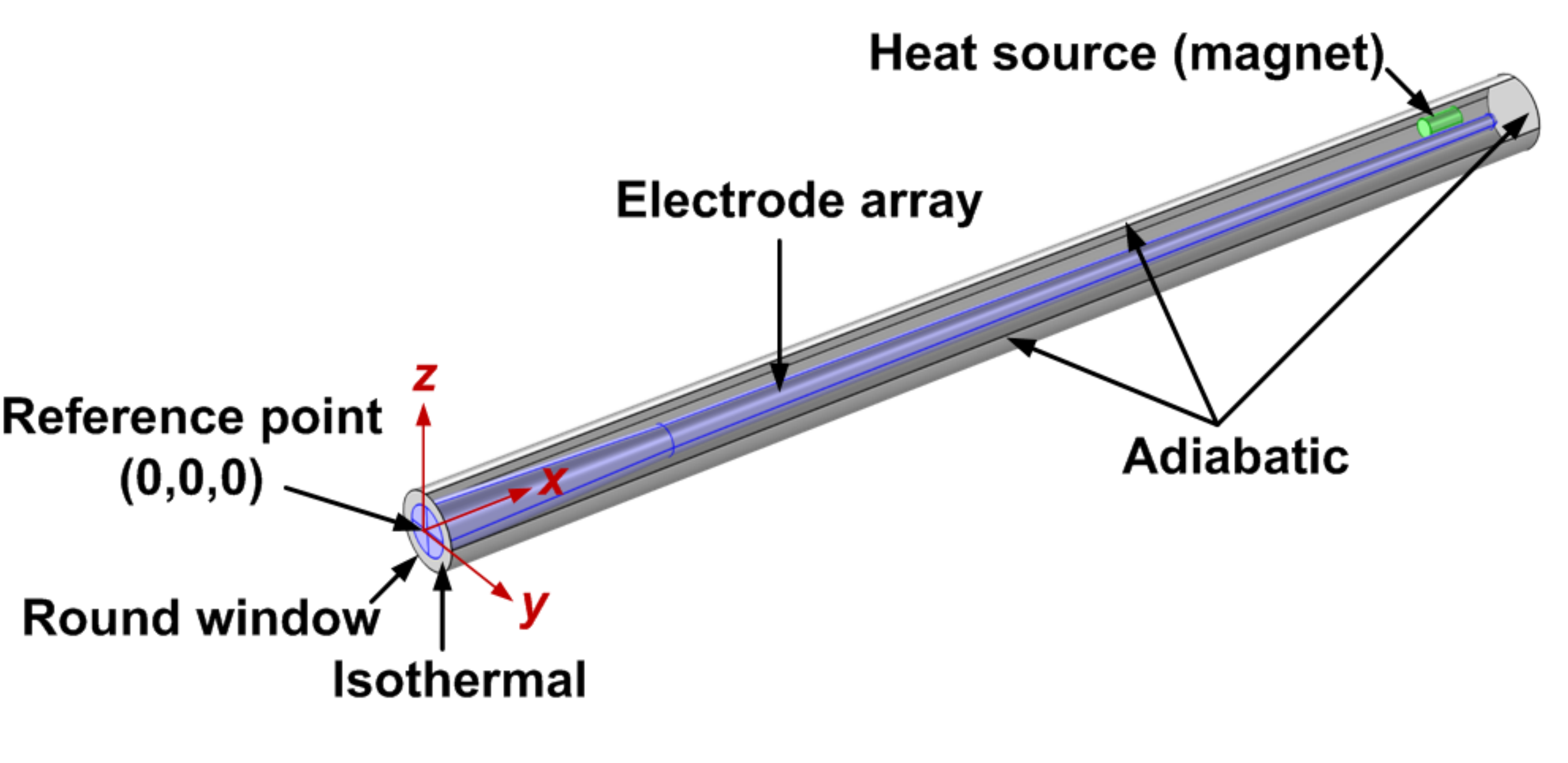}
\caption{\label{fig:2} 3D uncoiled model of the cochlea with inserted electrode array and magnet. The electrode array model is made by MED-EL \cite{DHANASINGH201793}.}
\end{figure}

The cochlea is a long semi-conical, spiral set of three fluid-filled ducts with two and one-half turns (see Fig. \ref{fig:1}). The fluid filling the cochlea is a dilute solution of ions in water called perilymph \cite{Perilymph}. In this paper, a 3D uncoiled model of the cochlea characterized by a length of 32.31 mm and a diameter of 2 mm is considered (see Fig.~\ref{fig:2}). A 31.5-mm-long electrode array is inserted in the cochlear canal through a dissected hole called the round window. The radius of the electrode array decreases linearly from 0.65 mm at the round window ($x$ = 0 mm) to 0.4 mm at $x$ = 6.5 mm, and then from 0.4 mm down to 0.2 mm between $x$ = 6.5 mm and 31.5 mm \cite{DHANASINGH201793}. The electrode array is surrounded by perilymph, and  the boundary of the cochlear canal is made of bone. A 1-mm-long, 0.5-mm-diameter cylindrical magnet acting as a heat source is aligned with and attached near the tip of the electrode array. The center of the magnet is located at $x$ = 30 mm and $y$= -0.7 mm.\\

Due to the relatively low maximum temperature involved in robotic cochlear implant surgery (a few degrees higher than the body core temperature of 37$^{\circ}$C), radiation heat transfer is negligible. As such, heat transfer in the 3D uncoiled model of the cochlea is analyzed by considering only conduction and natural convection heat transfer. The energy balance is given by:\\ 

\begin{equation} \label{eq:1}
{\rho c_p \frac{\partial T}{\partial t}} +{\rho c_p \mathbf{u} \cdot \nabla T} + {\nabla \cdot (-k \nabla T)} = q
\end{equation}\\
where $T$, $\rho$, $c_p$, $k$, $\mathbf{u}$, and $t$ are respectively temperature (K), density (kg/m$^3$), heat capacity (J/kg$\cdot$K), thermal conductivity (W/m$\cdot$K), velocity vector (m/s), and time (s).~The input power density, $q$ (W/m$^3$), is non-zero only in the magnet, while the advection term (i.e., second term on the left-hand side of Eq. (\ref{eq:1})) is non-zero only in the perilymph.~The magnet is heated by Joule heating.~The magnet detachment process starts after insertion of the cochlear implant and after switching off the external magnetic field. During insertion, the external magnetic field strength is on the order of 10 mT \cite{petruska}; thus, heating during this phase of the surgery is negligible. For contrast, a strong magnetic field of 3T, such as that produced by an MRI machine, only produces a temperature increase in a magnet of less than 0.5 $^{\circ}$C \cite{majdani}.  As such, the only heating within the cochlea that is considered occurs during the magnet detachment phase. The velocity field in the perilymph is determined by solving the following mass and momentum balance equations: 

\begin{equation} \label{eq:2}
\rho \frac{\partial \rho}{\partial t}+\nabla \cdot (\rho \bf{u})=0
\end{equation}
\begin{equation} \label{eq:3}
\rho \frac{\partial \bf{u}}{\partial t}+\rho(\bf{u}\cdot\nabla)\bf{u}=\nabla \cdot \it{p} \bf{I} + \rho\bf{g}
\end{equation}\\
where $\bf{g}$ (m/s$^{2}$) is the gravitational acceleration, ${p}$ is the pressure (Pa), and $\bf{I}$ is a $3\times3$ identity matrix. The magnetohydrodynamic effect is negligible in this analysis because the external magnetic field is removed after insertion of the magnet.\\

When solving the energy balance equation, the perilymph, electrode array, and magnet are initially at the body core temperature of 37$^{\circ}$C. Except at the round window where the bone is removed during surgery, the cochlear canal boundary is assumed to be insulated. This is justified by the fact that bones are characterized by a low thermal conductivity in the range of $\sim$0.373 to 0.496 W/m$\cdot$K \cite{bone}. At the round window, the perilymph and electrode array are isothermal at the body core temperature. Continuity boundary conditions are applied at the electrode array and magnet walls, which implies that the temperature and heat flux on these boundaries are equal for the adjacent domains.\\

For the mass and momentum balance equations, the perilymph is initially stagnant ($\bf{u}$ = 0 m/s), while the pressure inside the cochlear canal is equal to atmospheric pressure. The pressure at the round window is assumed to be constant and equal to atmospheric pressure. The magnet, cochlear canal, and electrode array walls are subjected to no-slip boundary conditions. The gravity effect is active along the negative $z$-direction and the reference point ($x=0$, $y=0$, $z=0$) is located at the round window in the center of the cochlear canal.\\

Equations (\ref{eq:1}) to (\ref{eq:3}) are solved using the finite element method as implemented in COMSOL Multiphysics 5.4
\cite{Bergheau}, \cite{reddy}, \cite{comsol} (see the Supplementary Material for computational details). In the calculations, the thermophysical properties of the perilymph are assumed to be the same as those of water \cite{Kassemi:2005},\cite{Perilymph}. The magnet and electrode array thermophysical properties are provided in Table~1. Before analyzing heat transfer in the cochlear canal shown in Fig.~\ref{fig:2}, the COMSOL model is first verified and validated. These are presented in the next section.\\

\begin{table}[h]
\begin{center}
\centering
\begin{threeparttable}
\label{thermalproperties}
\caption{Thermophysical properties of the magnet and electrode array.}
\begin{tabular}{l l l l} 
\hline
Domain & $c_{p}$~(J/kg$\cdot$K) & $k$~(W/m $\cdot$ K) & $\rho$~(kg/m$^3$)\\
\hline
Magnet   & 430 \tnote{a} & 8.1 \tnote{a} & 7500 \tnote{a}  \\
Electrode array & 127.7 \tnote{b} & 2.8 \tnote{b} & 19400 \tnote{b} \\
\hline
\end{tabular}
\begin{tablenotes}
\footnotesize
\item[a] Provided by the manufacturer (SUPERMAGNETMAN).
\item[b] Calculated based on the information provided by MED-EL.
\end{tablenotes}
\end{threeparttable}
\end{center}
\end{table}

\begin{figure}[h]
\centering\includegraphics[width=\linewidth]{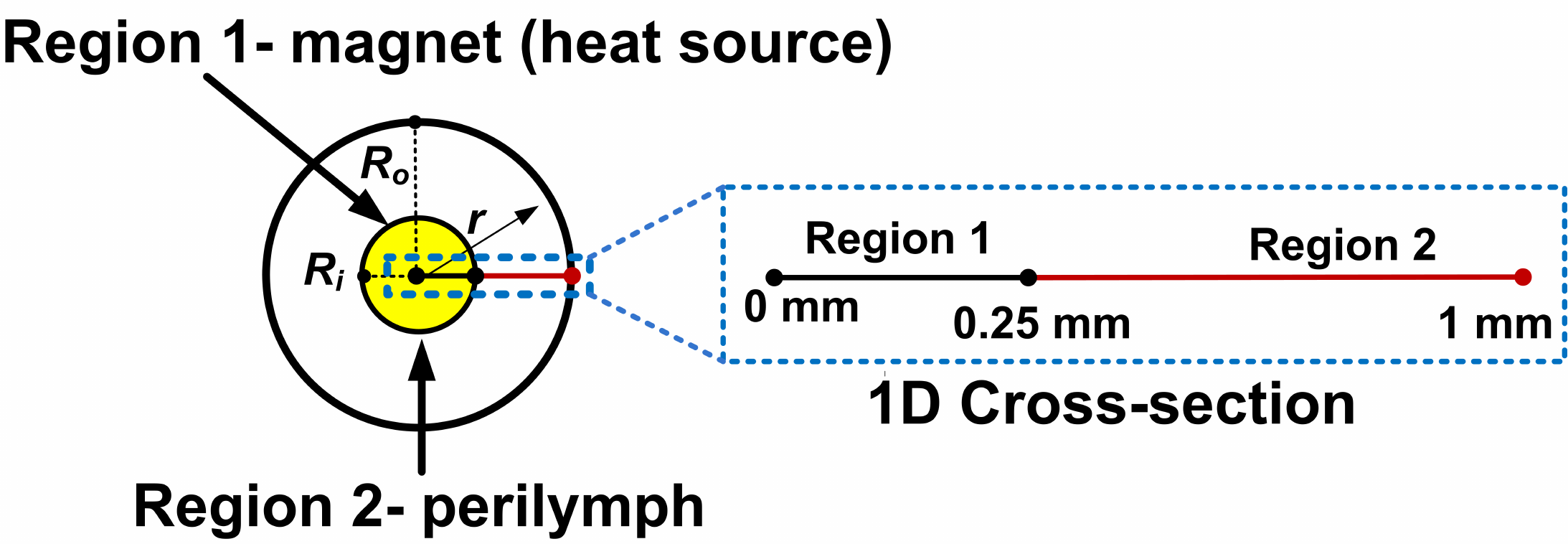}
\caption{\label{fig:3} The cochlear 3D model is uncoiled, and a cross-section of the uncoiled cochlea with inserted magnet represents the 1D model that is used for verification of conduction.}
\end{figure}


\section{Verification and validation of the model}
\subsection{\textbf{Verification of conduction heat transfer in a cross-section of the cochlea with heat source}}
Conduction heat transfer in the numerical model is verified in a 1D cross-section of the cochlea, where only temperature variations along the radial direction are analyzed (see Fig.~\ref{fig:3}).  The outer cylinder delimits the cochlear region, while the inner cylinder represents the magnet generating heat. The numerical results are compared with an analytical solution described hereafter.\\

The energy balances for pure conduction in the magnet (region 1: $r<R_i$ (see Fig. \ref{fig:3})) and in the perilymph (region 2: $R_i<r<R_o$ (see Fig. \ref{fig:3})) are respectively given by:\\
\begin{equation} \label{eq:4}
\alpha_{1}\frac{\partial^2 T_1}{\partial r^2}+\frac{\alpha_1}{k_1}q(r,t)=\frac{\partial T_1}{\partial t} 
\end{equation}
\begin{equation} \label{eq:5}
\alpha_{2}\frac{\partial^2 T_2}{\partial r^2}=\frac{\partial T_2}{\partial t} 
\end{equation}
where $\alpha$ and $r$ are the thermal diffusivity (m$^2$/s) and the radial coordinate (m). Subscripts 1 and 2 represent region 1 (magnet) and region 2 (perilymph), respectively. $R_i$ and $R_o$ are radii (m) of the inner and outer circles. 
At $r=R_i$, continuity boundary conditions are applied:
\begin{equation} \label{eq:6}
T_1(R_i,t)=T_2(R_i,t)
\end{equation}
\begin{equation} \label{eq:7}
k_1\left.\frac{\partial T_1}{\partial r}\right|_{r=R_i}= \left.k_2\frac{\partial T_2}{\partial r}\right|_{r=R_i}
\end{equation}
   
It is assumed that the entire domain is initially at the body core temperature ($T_{\textnormal{bc}}$ = 37$^{\circ}$C), while the input power density, $q$, is fixed at $3.3\times10^6$ W/m$^3$. The analytical solution for conduction heat transfer between two infinite, concentric cylinders is provided in Ref. \cite{ozisik}. Equations (\ref{eq:4}) and (\ref{eq:5}) are solved simultaneously using Green's functions.~The final result is a combination of Bessel functions of the first and second kinds that require computation of eigenvalues.~It is challenging, however, to calculate the first eigenvalue for an adiabatic cochlear boundary. Jain et al.~\cite{Jain} avoided using an insulation boundary condition, but no explanation was given. As such, a conductive boundary condition at $r=R_o$ is modeled with a small overall heat transfer coefficient, $U$, of 0.003 W/m$^2\cdot$K to mimic an adiabatic condition:

\begin{equation} \label{eq:8}
{-k_2\left.\frac{\partial T_2}{\partial r}\right|_{r=R_o}}+U(T_2(R_o) -T_{\textnormal{bc}})=0
\end{equation}

\begin{figure}[t]
\centering\includegraphics[width=4in]{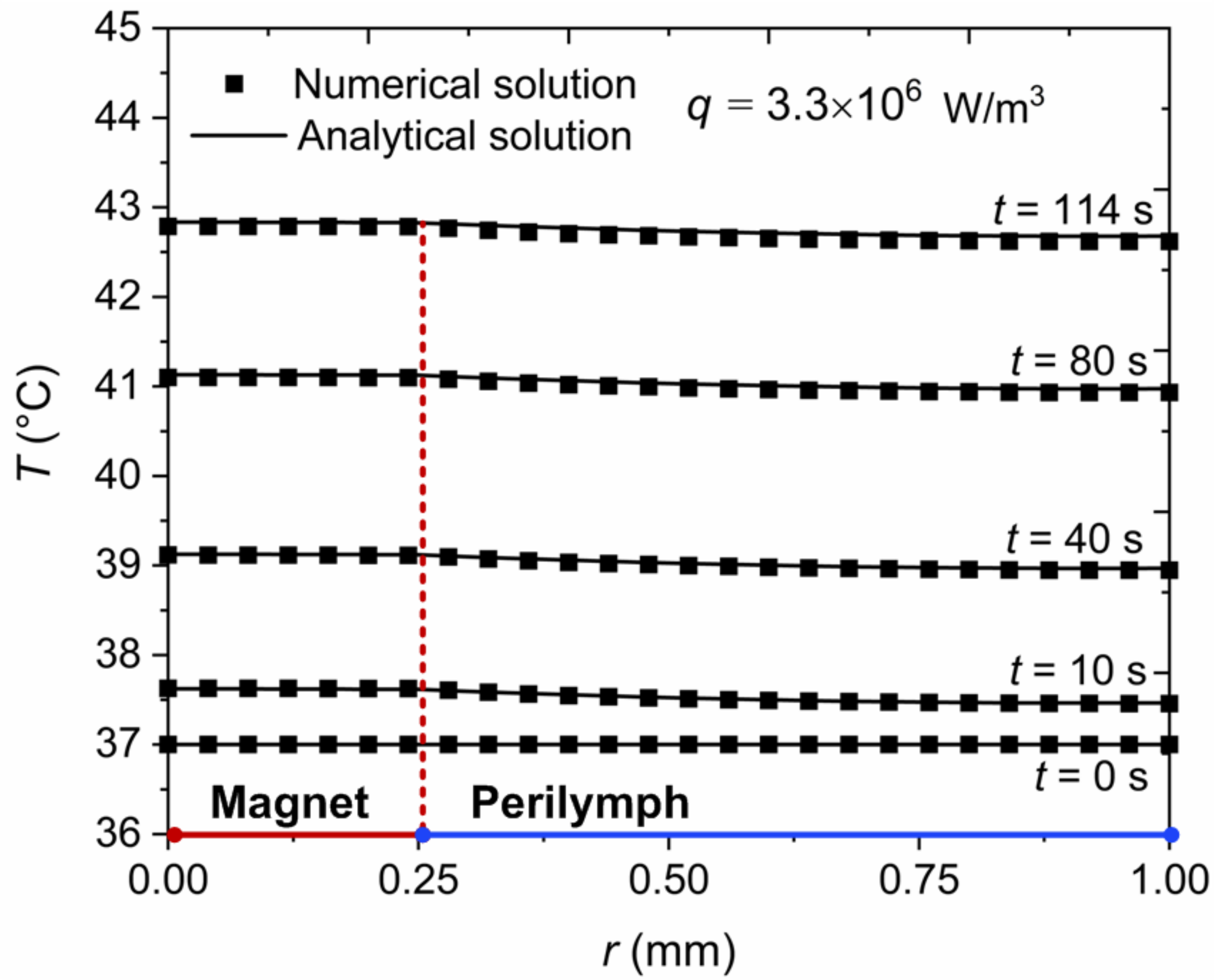}
\caption{\label{fig:4} Comparison of temperature profiles at selected times (0, 10, 40, 80, 114 s) from the analytical solution and the numerical model. The region between $r$=0 mm and 0.25 mm represents the magnet, while the rest of the domain is filled with perilymph.}
\end{figure}

 The convergence of the numerical solution has been studied by refining the element size as well as the time step. The numerical results converged using 8 elements and a time step of 0.1 s. The converged numerical results are verified against the analytical results in Fig.~\ref{fig:4}. The maximum Normal Root Mean Square Error (NRMSE) does not exceed 0.5\%. This difference is mostly due to truncation errors.\\ 
 
 Natural convection is the other heat transfer mode that may play a role in thermal energy dissipation during magnetic guidance of cochlear implants. Therefore, natural convection in the numerical model is verified and validated next using numerical and experimental data from the literature.\\

\subsection{\textbf{Verification and validation of natural convection in a cross-section of the cochlea with inserted magnet}}
\begin{figure}[tb]
\centering\includegraphics[width=4in]{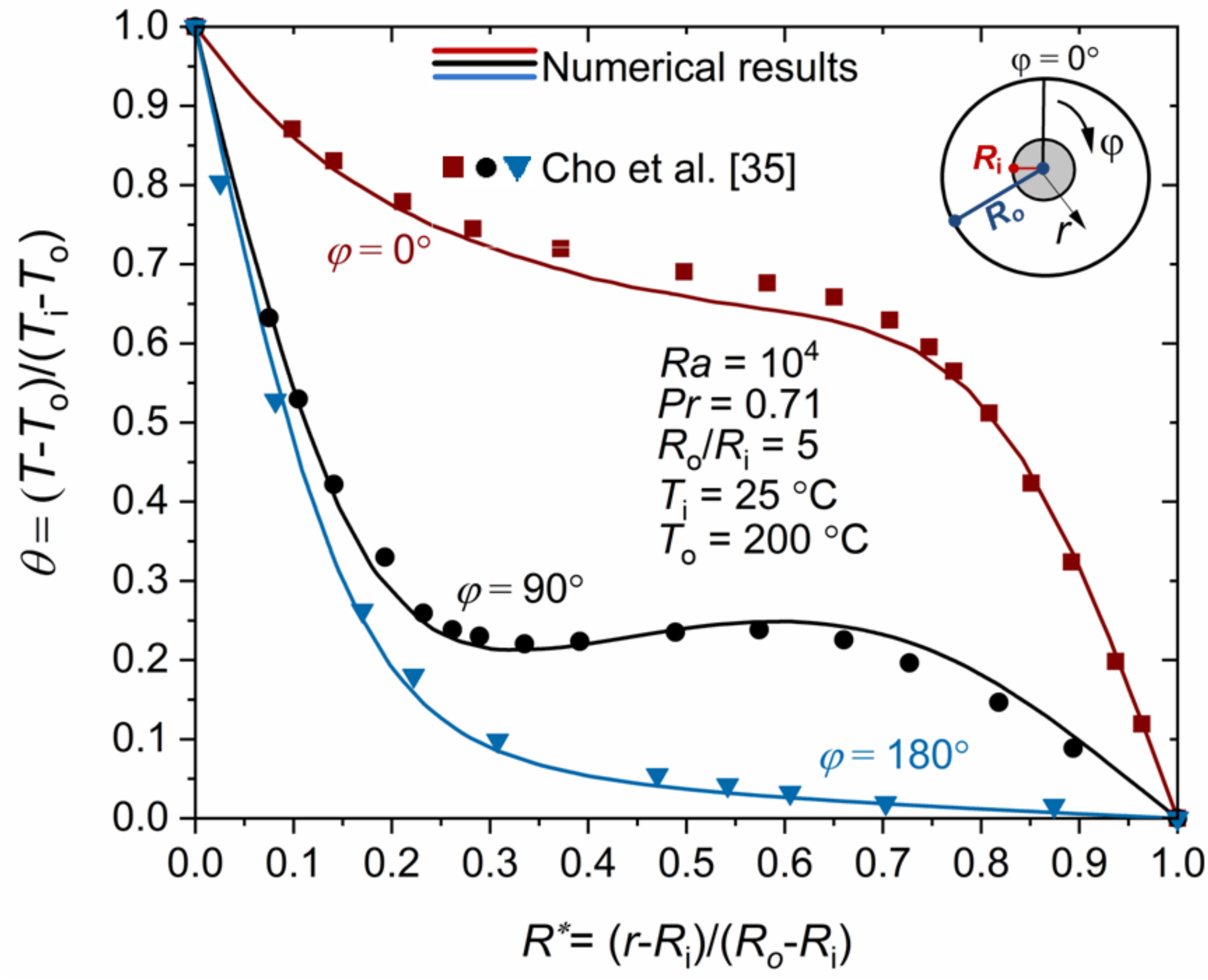}
\caption{\label{fig:5} Verification of natural convection in the numerical model with the results reported by Cho et al. \cite{cho} for two concentric cylinders. The vertical and horizontal axes represent the dimensionless temperature ($\theta$) and dimensionless radial distance ($R^{ \ast}$), respectively. }
\end{figure}

Natural convection in the numerical model is first verified with numerical data for two concentric, isothermal cylinders without heat generation. The energy, mass, and momentum equations given as Eqs. (\ref{eq:1}) to (\ref{eq:3}), respectively, are solved for an input power density ($q$) equal to zero. Radial temperature distributions from the numerical model are compared in Fig.~\ref{fig:5} against numerical results \cite{cho}  for selected azimuthal angles $\varphi$. The dimensional equations are solved and the results are nondimensionalized during the post-processing phase for easy comparison to temperature data presented in \cite{cho}. Here, $\theta$ is the dimensionless temperature $\frac{T-T_o}{T_i-T_o}$, where $T_o$ is temperature of the outer cylinder of radius $R_o$, while $T_i$ is the temperature of the inner cylinder of radius $R_i$. The dimensionless radial distance $R^*$ is defined as $\frac{r-R_i}{R_o-R_i}$. For the case depicted in Fig. \ref{fig:5}, the temperature difference between the inner and outer cylinders is 175 $^{\circ}$C, the Raleigh number $Ra = \frac{g \beta (T_i -T_o) (R_i-R_o)^3}{\nu \alpha}$ is $10^4$ which is within the range of natural convection, the Prandtl number $Pr$ is 0.71, and the ratio of the outer cylinder radius to the inner cylinder radius, $R_o/R_i$, is 5. \\

 A convergence analysis has been performed to determine the minimum number of elements leading to a stable solution. Initially,~142,961 elements were used to solve the problem; the number of elements was subsequently increased to~2,977,920. The maximum difference between these simulations was less than $1\%$. A time step of 0.1 s is used for the simulation. Refining the time step does not significantly affect the results. The maximum difference between the numerical results and those from Cho et al. \cite{cho} is less than 4$\%$. This difference may be due to using a digitizer tool (OriginPro 2019b) to extract the data from Ref. \cite{cho}, or the slight difference between the input parameters (e.g., material properties) used in our simulation and those from Cho et al \cite{cho}. \\ 
 
\begin{figure}[tb]
\centering\includegraphics[width=4in]{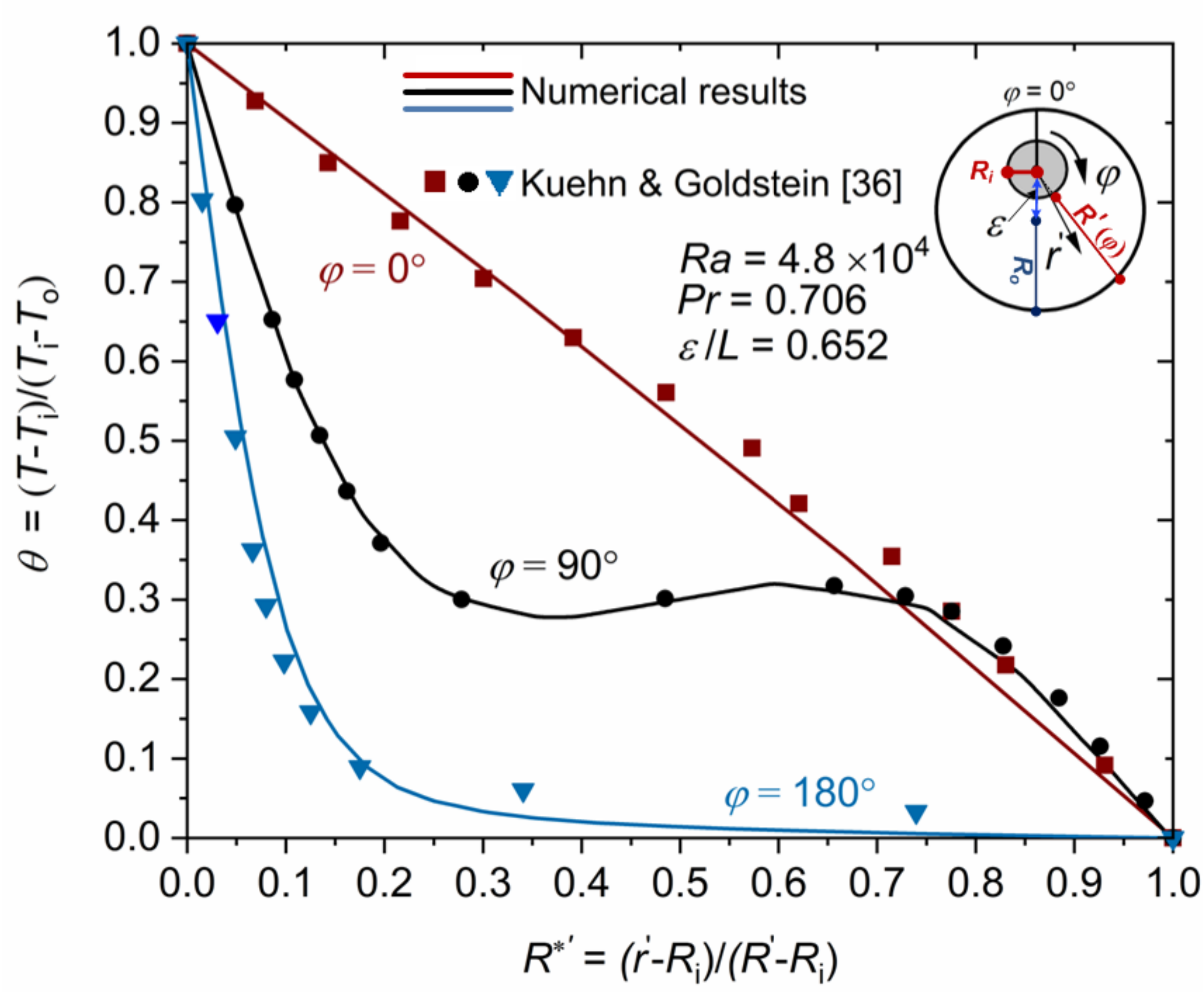}
\caption{\label{fig:6} Validation of natural convection in the numerical model with the results reported by Kuehn and Goldstein \cite{kuehn} for two eccentric cylinders. The vertical and horizontal axes represent the dimensionless temperature ($\theta$) and the dimensionless radial distance ($R^{ ^{*'}}$), respectively. }
\end{figure}

Natural convection is next validated against experimental data for two eccentric cylinders \cite{kuehn}, which is representative of the actual problem where the magnet attached to the electrode array is not centered in the cochlear canal (see Fig.~\ref{fig:6}). For this problem,  $r'$ represents the radial distance from the inner cylinder center, while $R'$ is the radial distance between the two cylinder walls. The ratio of the inner cylinder eccentricity, $\varepsilon$, to the gap distance between the two cylinders in a concentric arrangement, $(L = R_o-R_i)$, is  0.652. In addition, the Rayleigh number $Ra$ is $4.8 \times 10^4$, the Prandtl number $Pr$ is 0.706, while $R_i$  and  $R_o$ are respectively equal to 1.78 cm and 4.625 cm. The temperature difference, $\Delta T$, is 26.3 $^{\circ}$C, and $T_i+T_o$~=~70.46 $^{\circ}$C.\\

A convergence analysis of the numerical model has been performed in the same manner as for two concentric cylinders.~The number of elements leading to converged results is 205,021.~At time steps shorter than 0.1 s, the numerical results do not change by more than 1$\%$.~The dimensionless temperature ($\theta$) is plotted as a function of the dimensionless radial the distance $R^{*'}$=$\frac{r'-R_i}{R'-R_i}$ for selected values of $\varphi$ in Fig.~\ref{fig:6}. The maximum NRMSE  is 6$\%$, which is equal to 1.7 $^{\circ}$C. The error in digitizing the data from Ref. \cite{kuehn}, and the measurement error (not specified explicitly) are the main plausible sources of differences.\\ 
 
To conclude this section, the numerical model provides accurate results for both conduction and natural convection heat transfer. As such, the numerical model can be applied with confidence to the thermal analysis of the uncoiled cochlea shown in Fig.~\ref{fig:2}. 

\section{Heat transfer analysis in the 3D uncoiled model of the cochlea with inserted electrode array and magnet}

The verified and validated numerical model is used to simulate heat transfer in the 3D uncoiled model of the cochlea with an inserted electrode array and magnet, as described in section 2 and shown in Fig.~\ref{fig:2}. In all simulations, 3,565,495 elements and time steps of 0.1 s were sufficient to obtain converged results. Note that the number of elements is reduced to 808,937 when natural convection is neglected. Prior to discussing the heat transfer analysis, the computational details are presented in subsections 4.1 and 4.2 (see Supplementary Material for details on element distribution, and solver settings).

\subsection{\textbf{Grid independence}}
\begin{figure}[h]
\centering\includegraphics[width=\linewidth]{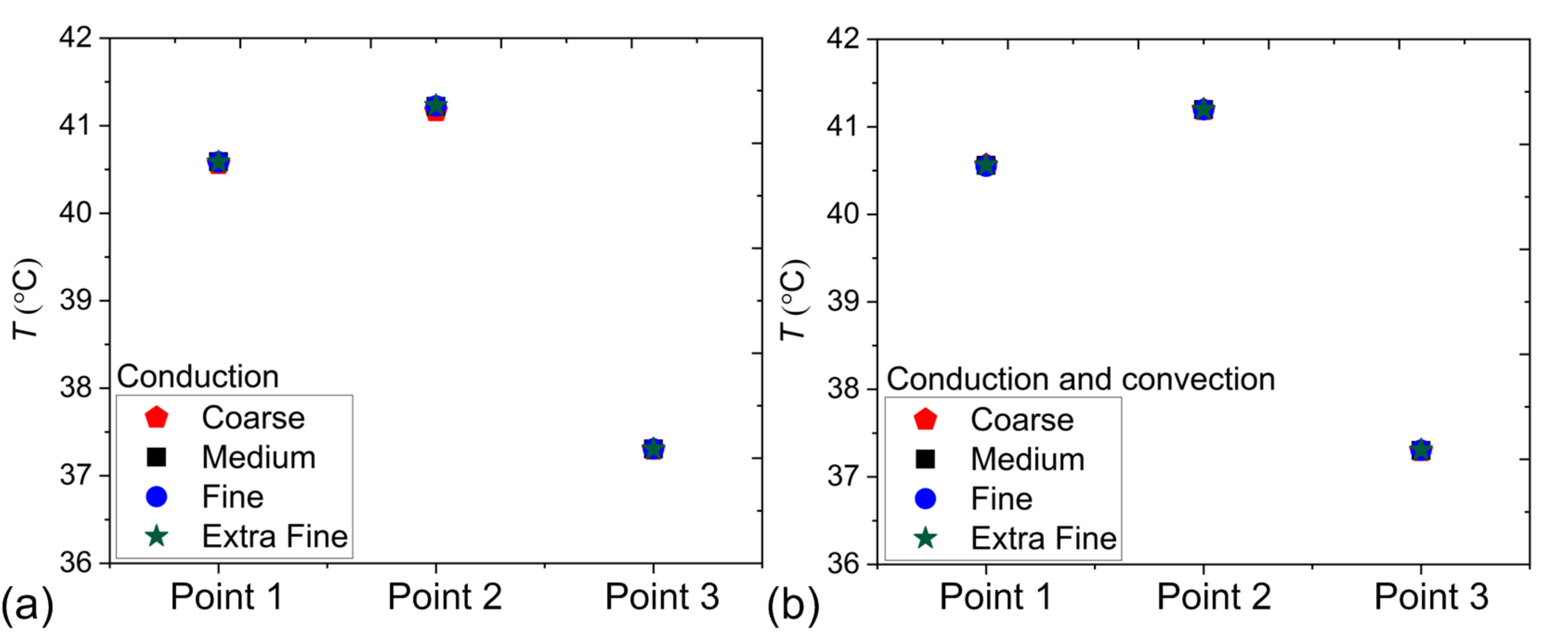}
\caption{\label{gridindependence} Grid independence results in three points inside the cochlea for four different grid sizes ($q$ = 1.62 $\times$ 10$^7$ W/m$^3$, $t$ = 114 s): (a) conduction without convection, (b) conduction with convection.}
\end{figure}

Considering conduction only, the number of elements has been increased in four steps to determine the minimum number of elements necessary such that the solution does not change with the mesh size. The coarse, medium, fine, and extra fine meshes include 18,067, 110,087, 304,362, and 808,937 (808,937 Tetrahedra, 51,380 Triangles, 3,028 Edge elements, and 32 Vertex elements)  elements, respectively. The temperatures at $t$ = 114 s in three locations in the cochlea are selected to assess grid independence (see Fig. \ref{gridindependence} (a)). Point 1 (30.5 mm, -0.96 mm, 0.0 mm) is located on the left side of the magnet, point 2 (30.5 mm, -0.4 mm, 0.0 mm) on the right side of the magnet, and point 3 (15 mm, -0.6 mm, 0.0 mm) is located in the middle of the cochlea on the left side of the electrode array. The maximum difference between the temperature data is 0.9$\%$.

In the case of conduction with convection, the coarse, medium, fine, and extra fine meshes include 224,191, 429,652, 1,093,677,and 3,565,495 (3,214,283 Tetrahedra, 592 Pyramids, 350,620 Prisms, 177,936 Triangles, 344 Quads, 4,816 Edge elements, and 32 Vertex elements) elements, respectively. Once again, the temperatures at $t$ = 114 s in the same three locations in the cochlea are selected to determine grid independence (see Fig. \ref{gridindependence} (b)). The maximum difference between the temperature data is 0.09$\%$. Note that $y^+$ is less than 1 indicating that there is sufficient mesh resolution near the wall. The Rayleigh number $Ra$ is less than $10^3$, thus the flow is laminar. The grid independence presentation in this document is based on Refs. \cite{reviewer-paper1}, \cite{reviewer-paper2}, \cite{reviewer-paper3}.

\subsection{\textbf{Time step analysis}}
The finest meshes with 808,937 elements for conduction and 3,916,115 elements for conduction and convection are used to determine the impact of the time step on the final results. The three values of time step selected for this study are: 0.05 s, 0.1 s, and 0.5 s. Temperature data at the same three locations used to study grid independence are employed here. The results are provided in Tables \ref{timedeanalysis1} and \ref{timedeanalysis2}. The time step does not have any impact on the conduction results. In the case of conduction with convection, the data change by less than 0.1 $\%$.
\begin{table}[h]
\caption{Calculated temperature ($q$ = 1.62 $\times$ 10$^7$ W/m$^{3}$, $t$ = 114 s ) at three locations for simulations of conduction without convection as a function of time step.}
\begin{center}
\centering
\small
\label{timedeanalysis1}
\begin{tabular}{l l l l} 
\hline
Time Step (s) & $T_{\rm{point1}}$~($^{\circ}$C) & $T_{\rm{point2}}$~($^{\circ}$C) & $T_{\rm{point3}}$~($^{\circ}$C)\\
\hline
0.05 & 40.584 & 41.226 & 37.306\\
0.1 & 40.584 & 41.226 & 37.306\\
0.5 & 40.584 & 41.226 & 37.306\\
 
\hline
\end{tabular}
\end{center}
\end{table}
\begin{table}[h]
\caption{Calculated temperature ($q$ = 1.62 $\times$ 10$^7$ W/m$^3$, $t$ = 114 s ) at three locations for simulations of conduction with convection as a function of time step.}
\begin{center}
\centering
\small
\label{timedeanalysis2}
\begin{tabular}{l l l l} 
\hline
Time Step (s) & $T_{\rm{point1}}$~($^{\circ}$C) & $T_{\rm{point2}}$~($^{\circ}$C) & $T_{\rm{point3}}$~($^{\circ}$C)\\
\hline
0.05 & 40.561 & 41.201 & 37.308\\
0.1 & 40.563 & 41.200 & 37.306\\
0.5 & 40.562 & 41.201 & 37.308\\
 
\hline
\end{tabular}
\end{center}
\end{table}
\subsection{\textbf{Impact of natural convection on heat transfer within the cochlea}}
The thermal damage threshold of tissues in the cochlea is required for calculating the maximum safe input power density to detach the magnet  \cite{thermaldose1}. This thermal damage threshold of in-vivo tissues depends on the temperature, the type of tissues, and the length of exposure to the heat source. The Cumulative Equivalent Minutes at a fixed temperature ($CEM_{T}$) is a parameter that combines both the effects of temperature and length of exposure \cite{thermaldose1}, \cite{Yoshida10116}. As such, $CEM_{T}$ is used to determine the maximum safe input power density for detaching the magnet from the electrode array. Yoshida et al. \cite{Yoshida10116} reported that exposing mouse ear tissues to a temperature of 43$^{\circ}$C for 1.9 minutes (114 s) does not affect ear functionality. Van Rhoon et al. \cite{Rhoon} pointed out that a $CEM_{43}$ less than 2 minutes is safe for any type of tissues under supervision of an expert capable of managing a sudden physiological response to a thermal stress. Here, a $CEM_{43}$ of 144 s is used to calculate the maximum safe input power density.\\
 
The maximum safe allowable input power density for 114 s of heating is first estimated for the limiting case of pure conduction within a cochlea containing a solitary magnet. This is the worst case scenario, as natural convection and conduction through the electrode array facilitate heat dissipation in the cochlea. By applying the Parametric Sweep Study Module in COMSOL, it is found that the maximum safe input power density for this limiting case is $1.62\times10^7$ $\textup W/ \textup m^3$ based on a $CEM_{43}$ of 114 s. This input power density is used hereafter to study the impacts of natural convection and conduction through the electrode array on the thermal management of the cochlea. \\
 
\begin{figure}[h]
\centering\includegraphics[width=\linewidth]{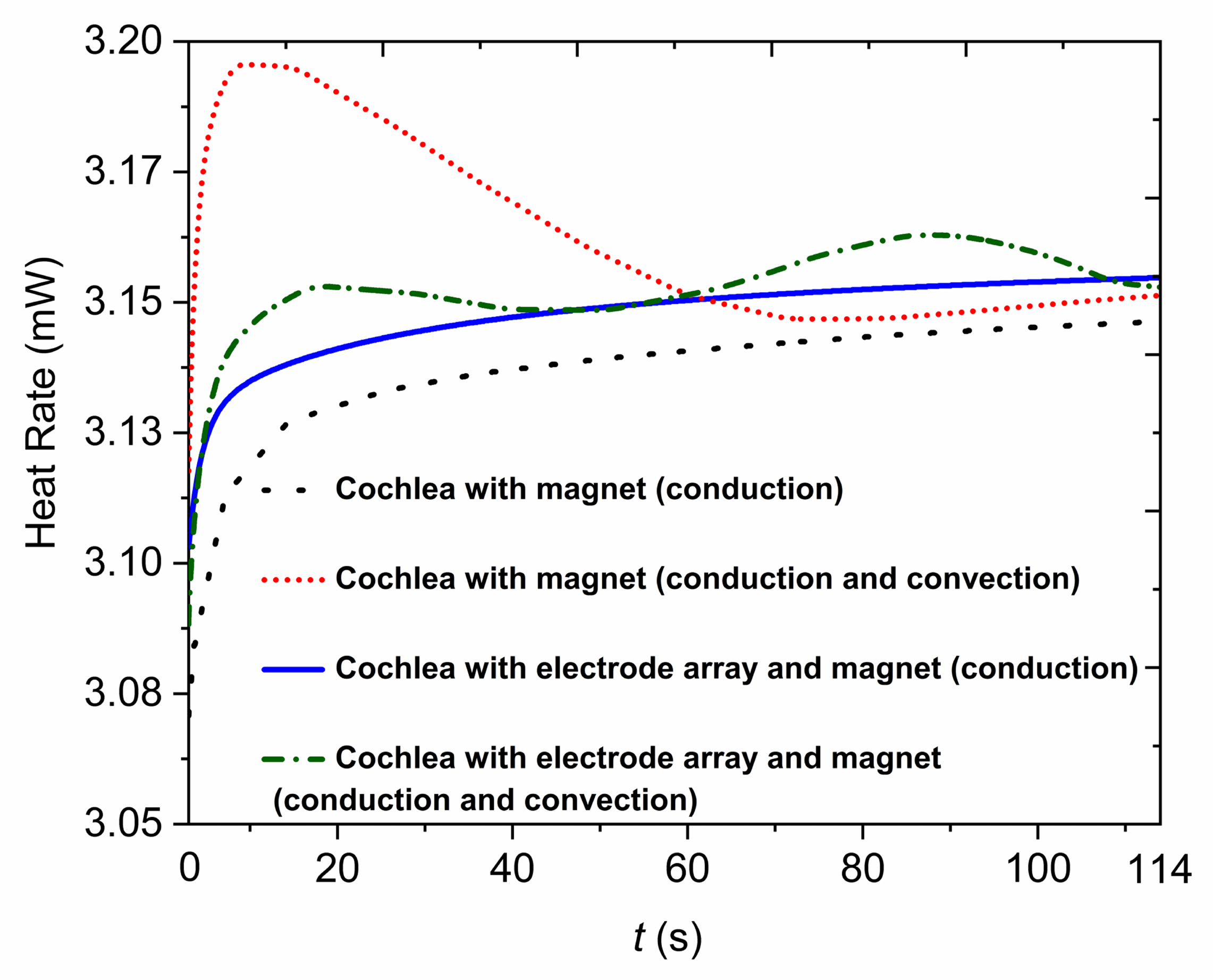}
\caption{\label{heatremoval} Heat removal rate from the magnet as a function of time.}
\end{figure}

\begin{figure}[h]
\centering\includegraphics[width=\linewidth]{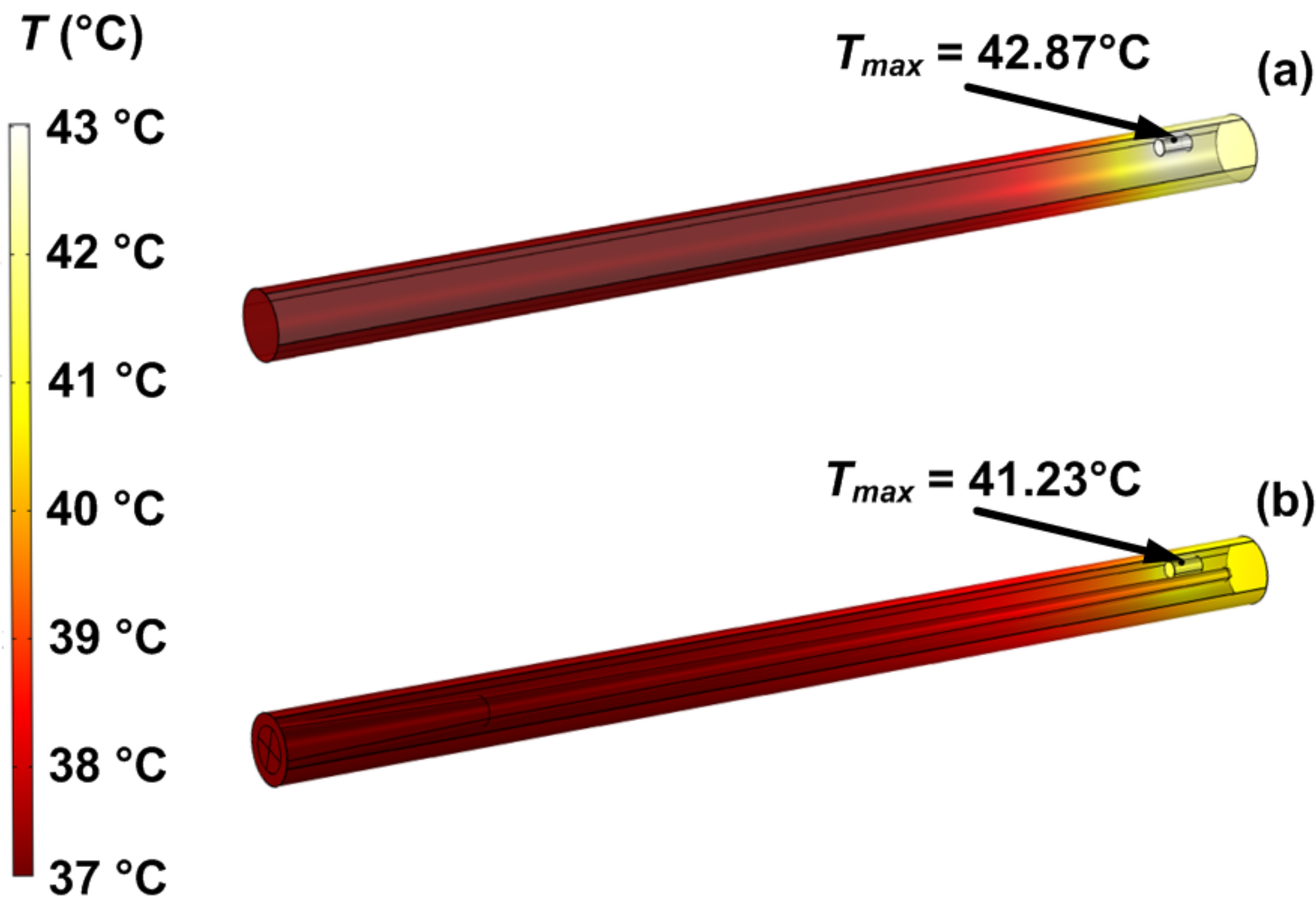}
\caption{\label{tempprofile} Temperature distribution at $t = 114$ s in the uncoiled cochlea due to heat transfer by conduction: (a) With a solitary magnet. (b) With electrode array and magnet.}
\end{figure}

The maximum temperature in the cochlea is provided in Table~\ref{table4} for four different scenarios, namely for pure conduction with and without an electrode array, and for conduction and natural convection with and without an electrode array. In the absence of the electrode array, natural convection reduces the maximum temperature in the cochlea by approximately 1$^\circ$C. This effect is significant considering the fact that an increase of temperature in excess of 6$^\circ$C above the body core temperature causes damage to tissues. Conversely, in the cochlea with electrode array, the maximum $Ra$ is less than 400, so  the flow is laminar and conduction is dominant \cite{incropera}. Also, the maximum P\'{e}clet number ($Pe$) is 0.33, which means that the diffusion transport rate dominates the advection transport rate. Consequently, the impact of natural convection on the maximum temperature in the cochlea with the electrode array is negligible. Thus, it can be concluded that inserting the electrode array, as done in the actual surgery, reduces the relative contribution of natural convection to heat transfer in the cochlea.~This conclusion is also confirmed in Fig.~\ref{heatremoval}, where the heat removal rate from the magnet is shown as a function of time.~When the electrode array is not modeled, the heat rate from the magnet increases in a non-negligible manner due to natural convection during the first 60 s of the transient process.~Yet, the difference between the heat rate removal from the magnet by natural convection is clearly negligible in comparison to the heat rate removal by conduction in the presence of the electrode array. These results can be explained by the fact that the electrode array drives some perilymph out of the cochlea. The remaining fluid in the small annular region does not flow easily due to the internal no-slip boundary condition around the electrode array. As such, heat is mostly transferred axially via conduction in the electrode array. Fig.~\ref{tempprofile} provides the temperature distribution within the uncoiled cochlea with and without the electrode array after heating the magnet for 114 s with an input power density of $1.62\times10^7$ W/m$^3$. Perilymph temperature is maximum near the magnet and reduces to the body core temperature when approaching the round window. \\

The negligible impact of natural convection in the presence of the electrode array is further supported by the correlation developed by Raithby and Holland \cite{RAITHBY} for two concentric cylinders separated by a fluid gap assumed to be much smaller than the length of the cylinders.~In this correlation, an effective thermal conductivity $k_{\textnormal{eff}}$ (W/m$\cdot$K) due to conduction and natural convection within the gap between two isothermal cylinders is calculated as follows:
\begin{equation} \label{eq:9}
\frac{k_{\textnormal{eff}}}{k}= 0.386\left(\frac{Pr}{0.861+Pr}\right)^{\frac{1}{4}}Ra_{cc}^{\frac{1}{4}}
\end{equation}\\
where $k$ is the thermal conductivity of the fluid in the gap. The correlation assumes that the temperature of the inner cylinder is greater than the outer cylinder temperature. In Eq.~(\ref{eq:9}), the modified Rayleigh number for two concentric cylinders, $Ra_{cc}$, is defined as \cite{RAITHBY}:
\begin{equation} \label{eq:10} Ra_{cc}=\frac{[ln(\frac{D_o}{D_i})]^{4}}{b^3(D_i^{-3/5}+D_o^{-3/5})^{5}}Ra_b
\end{equation}\\
where $b=D_o-D_i$, and $D_o$ and $D_i$ are respectively the outer and inner cylinder diameters. Equations (\ref{eq:9}) and (\ref{eq:10}) are applicable for fluids characterized by $0.7\leq Pr \leq 6000$ and $Ra \leq 10^7$. A ratio $\frac{k_\textnormal{eff}}{k}$ larger than 1 indicates that natural convection contributes to heat transfer in a non-negligible manner. Otherwise, natural convection is negligible and heat transfer can solely be modeled via conduction. Substituting the cochlear dimensions into Eqs. (\ref{eq:9}) and (\ref{eq:10}), and using the temperature difference $\Delta{T}$ = 43$^{\circ}$C - 37$^{\circ}$C to calculate $Ra_b$, the maximum $\frac{k_{\textnormal{eff}}}{k}$ ratio is 0.7. Therefore, both the numerical model and correlation demonstrate that natural convection is negligible in the thermal analysis of the magnet detachment process. The maximum safe input power for a range of heating time intervals is discussed in the next section.

\begin{table}[h]
\caption{Maximum temperature in a cochlear canal.}
\begin{center}
\centering
\begin{adjustbox}{width=1\textwidth}
\small
\label{table4}
\begin{tabular}{l l} 
\hline
 & $T_{max}$~($^{\circ}$C) \\
\hline
Cochlea with magnet (conduction) & 42.87  \\
Cochlea with magnet (conduction and convection) & 42.08  \\
Cochlea with electrode array and magnet (conduction) & 41.23  \\
 Cochlea with electrode array and magnet (conduction and convection) & 41.20  \\
\hline
\end{tabular}
\end{adjustbox}
\end{center}
\end{table}

\section{Determination of the safe input power density range}
\begin{figure}[h]
\centering\includegraphics[width=\linewidth]{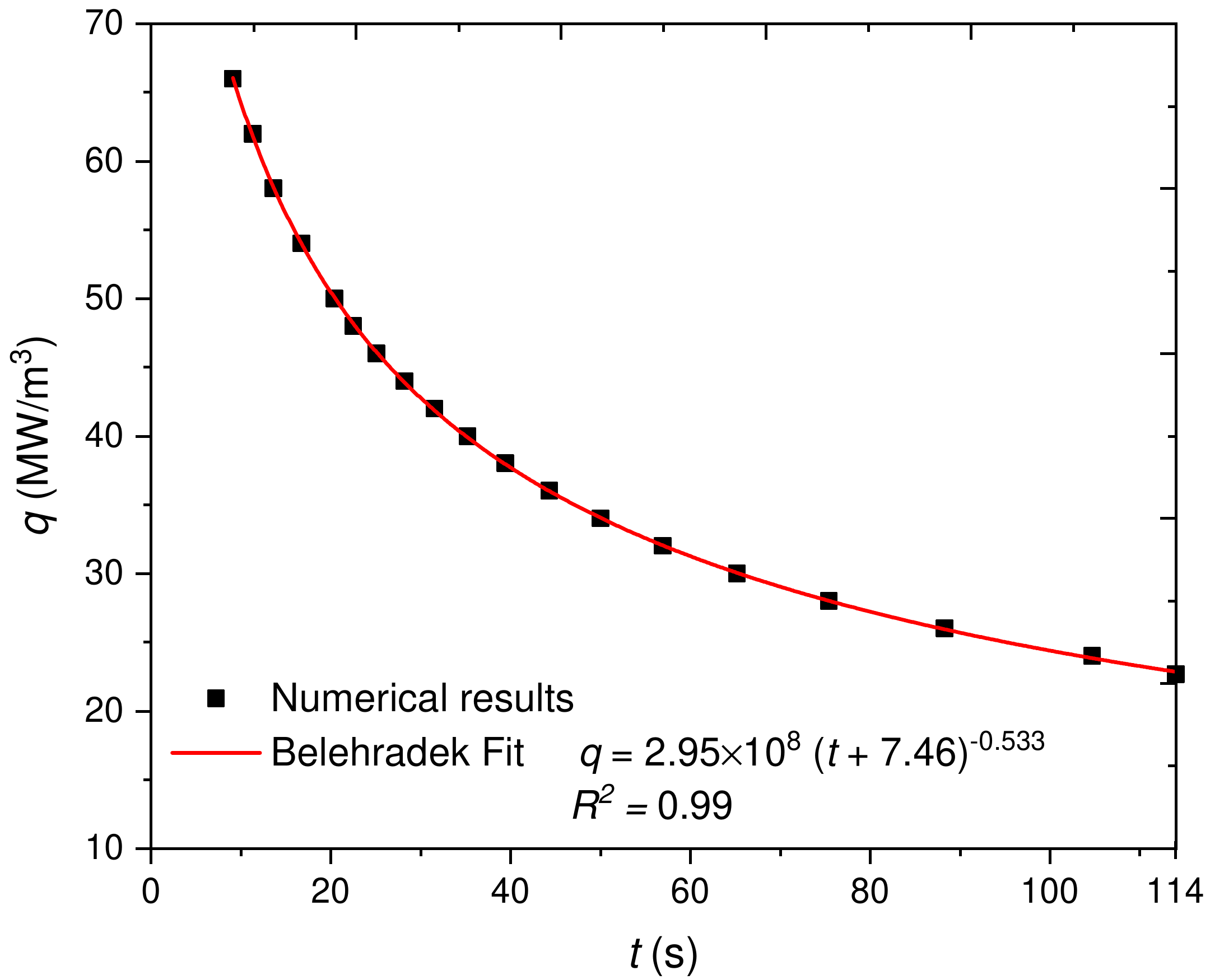}
\caption{\label{maxpower} Maximum safe input power density as a function of heating time interval.}
\end{figure}

The range of safe input power is determined by limiting the maximum allowable temperature within the entire cochlea to 43$^{\circ}$C. Thus, the safe input power density depends on the length of the heating period, i.e., faster detachment requires higher input power density. A parametric study is performed to determine the maximum safe input power density at discrete time intervals (see Fig. \ref{maxpower}). By fitting the data to a power law function, the maximum input power density is shown to be inversely proportional to approximately the square root of the heating period.  Total energy transferred to the magnet decreases by increasing the input power density and concurrently reducing the heating period. In a faster heating process, the rate of input energy is higher than the rate of the heat removal by perilymph in comparison to the slower heating case. Consequently, the temperature of the magnet and the immediate adjacent region ascends quickly, as a result, lower energy can be transferred to the magnet to prevent thermal trauma.\\

One possible means to attach the magnet to the tip of the electrode array is by  a paraffin wax structure. To determine the energy required to melt the paraffin, it is assumed that the paraffin is a lumped capacitance, all the energy inside the magnet is transferred to the paraffin, and the paraffin is not exchanging heat with the surrounding material. With these assumptions, the combined sensible heat and the latent heat required to melt a 0.5 mm$^{3}$ paraffin bulk with a melting point of 43$^{\circ}$C is approximately 0.1 J~(1.8  $\times$ 10$^{6}$~{W/m$^{3}$} for a heating duration of 114 s), which is about one order of magnitude lower than the maximum safe input power. The lowest maximum safe input power is sufficient to melt the paraffin around the magnet to release it. Therefore, a paraffin structure can be used to attach the magnet, and after the insertion, the paraffin can be melted without the risk of hyperthermia. 
\section{Impact of the cochlear fluid on transient temperature change}
\begin{figure}[h]
\centering\includegraphics[width=\linewidth]{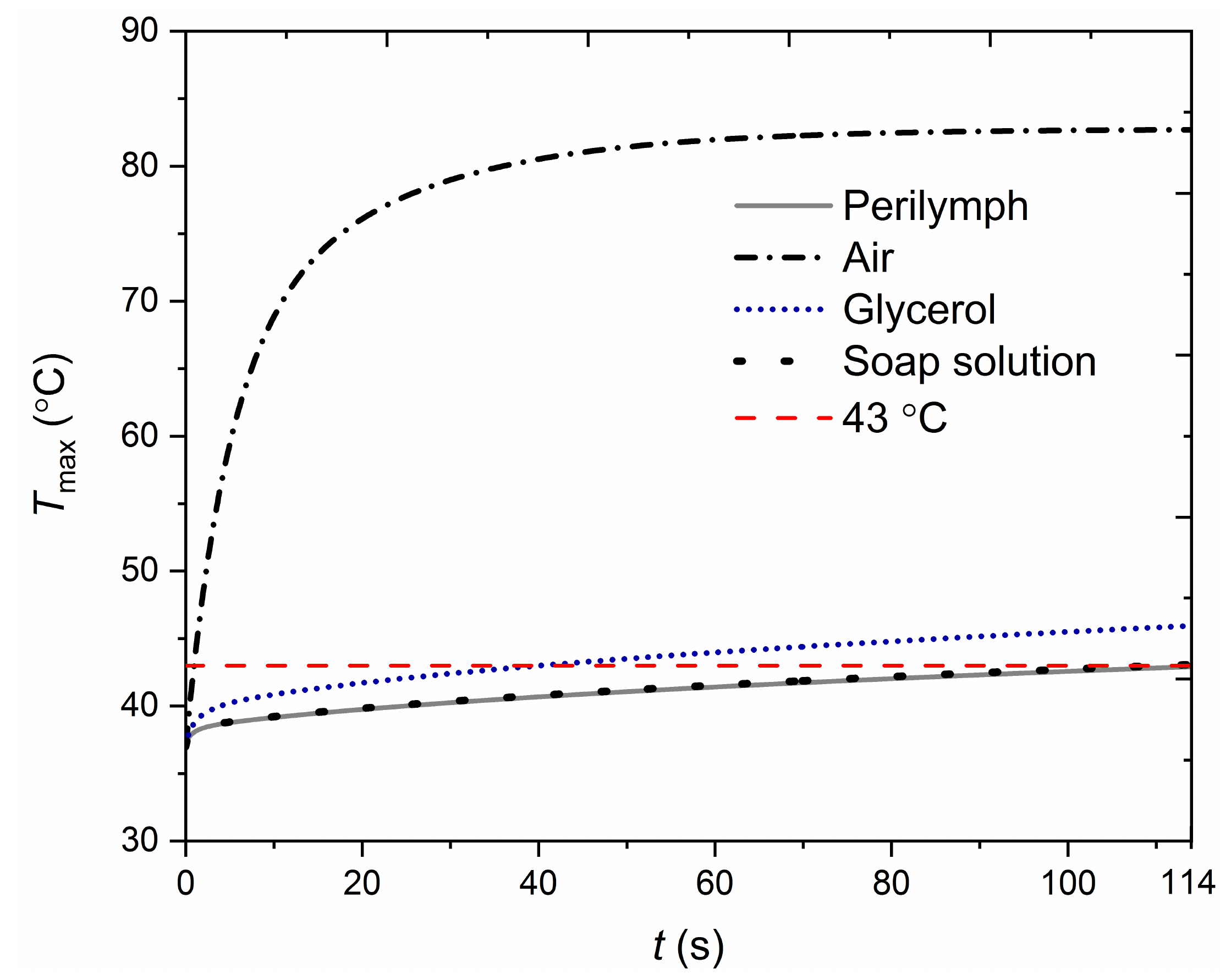}
\caption{\label{materialimpact} Transient maximum temperature within the uncoiled cochlea with inserted magnet and electrode array filled with perilymph, air, Glycerol, and soap solution ($q$ = 2.265$\times$10$^7$ W/m$^3$ for 114 s).}
\end{figure}

The fluid within the cochlea is an important parameter to determine the safe input power density for magnet detachment. During cochlear implant electrode array insertion, some amount of perilymph is forced from the cochlea and is replaced by air. This scenario represents a worst case scenario in terms of possible exposure to hyperthermia. Thus, the impact of replacing the perilymph with air should be studied. Research also suggests that lubricants can reduce intracochlear force during insertion \cite{lubricant}. Employing a lubricant during magnetic insertion also has potential for decreasing the intracochlear physical trauma. Thus, the effect of the various fluids on the magnet detachment process is critical. Four fluids, including perilymph, air, Glycerol, and a soap solution (10$\%$ wt soap, 90$\%$ wt distilled water \cite{lubricant}) are studied to determine their impact on temperature increase within the cochlea during detachment (see Fig. \ref{materialimpact}). Thermophysical properties of air, perilymph, Glycerol, and the soap solution are provided in Table~5. Air has the largest thermal diffusivity among these four substances, and consequently the rate of temperature change for air exceeds that of the other three fluids. The data from this study indicate  that the cochlea should be filled with either perilymph or the soap solution, otherwise, the magnet (and surrounding tissues) will heat too quickly for safe control or without significantly reducing the input power density.~The soap solution decreases the insertion force and is thermally compatible with the magnetic insertion of the electrode array since the primary component is distilled water. As a result, conducting magnetic insertion with the soap solution has potential for future use.
\begin{table}[h]
\begin{center}
\centering
\begin{threeparttable}
\label{table5}
\caption{Thermophysical properties of four possible cochlear fluids.}
\begin{tabular}{l l l l l} 
\hline
Fluid & $c_{p}$~(J/kg$\cdot$K) & $k$~(W/m$\cdot$K) & $\rho$~(kg/m$^3$)& $\alpha$(m$^2$/s)\\
\hline
Air  & 1006.4 \cite{comsol} & 0.027 \cite{comsol} & 1.13 \cite{comsol} & 2.37$\times 10^{-5}$\\
Perilymph  & 4176.6 \cite{comsol} & 0.625 \cite{comsol} & 992.20 \cite{comsol} & 1.51$\times 10^{-7}$\\
Glycerol & 2240.0 \cite{glycerol} & 0.285 \cite{glycerol} & 1260.00 \cite{glycerolcp} & 1.01$\times 10^{-7}$\\
Soap solution  & 4000.0 \cite{soap} & 0.600 \cite{soap1} & 999.00 \cite{soap} & 1.50$\times 10^{-7}$\\
\hline
\end{tabular}

\footnotesize

\end{threeparttable}
\end{center}
\end{table}
\section{Conclusions}
A 3D uncoiled model of the cochlea was, for the first time, developed to analyze heat transfer within cochlear canals. Specifically, the range of safe input power density to detach the magnet from the electrode array during robotic cochlear implant surgery was studied using a 3D uncoiled model of the cochlea. Conservation of mass, momentum, and energy in the cochlea were solved using the finite element method as implemented in COMSOL Multiphysics 5.4. The numerical model was verified and validated for conduction and natural convection heat transfer. It was found that natural convection has a negligible impact on dissipating the heat generated during the magnet detachment process as most of the heat is transferred axially by conduction through the electrode array. Solving the equations for conservation of mass, momentum, and energy simultaneously, which is required to calculate natural convection, is computationally expensive. Thus, the fact that natural convection is negligible is critical for modeling heat transfer in the actual cochlea geometry as it reduces the computational time drastically from about 32 hrs for conduction and natural convection to 1 hr for conduction only on a desktop computer with an Intel Core i9-9900K  processor and 64.0 GB RAM. Finally, the range of safe maximum input power density to detach the magnet after magnetic guidance of the cochlear implant is 2.265$\times$10$^7$ W/m$^3$ for 114s to 6.6$\times$10$^7$ W/m$^3$ for 9s. A soap solution is a suitable lubricant that decreases insertion forces and is safe to  use during the magnet detachment process. This work will accelerate the implementation of robotic implant cochlear surgery and is critical for avoiding thermal damage of cochlear tissues. In addition, this research provides a foundation for heat transfer studies of a coiled cochlea. \\

\section*{Acknowledgements} 
Research reported in this publication was supported by the National Institute on Deafness and Other Communication Disorders of the National Institutes of Health under Award Number R01DC013168. The content is solely the responsibility of the authors and does not necessarily represent the official views of the National Institutes of Health. We acknowledge the Center for High Performance Computing at the University of Utah for their support and resources. We would also like to show our gratitude to the MED-EL company for sharing the information about their electrode arrays.


\bibliography{02-References.bib}

\begin{thebibliography}{10}
\expandafter\ifx\csname url\endcsname\relax
  \def\url#1{\texttt{#1}}\fi
\expandafter\ifx\csname urlprefix\endcsname\relax\def\urlprefix{URL }\fi
\expandafter\ifx\csname href\endcsname\relax
  \def\href#1#2{#2} \def\path#1{#1}\fi

\bibitem{Clark}
J.~Clark, L.~Leon, F.~Warren, J.~Abbott, Magnetic guidance of cochlear
  implants: Proof-of-concept and initial feasibility study, Journal of Medical
  Devices, Transactions of the ASME 6~(3).
\newblock \href {http://dx.doi.org/10.1115/1.4007099}
  {\path{doi:10.1115/1.4007099}}.

\bibitem{Clark2}
J.~R.~Clark, F.~Warren, J.~Abbott, A scalable model for human scala-tympani
  phantoms, Journal of Medical Devices 5.
\newblock \href {http://dx.doi.org/10.1115/1.4002932}
  {\path{doi:10.1115/1.4002932}}.

\bibitem{Leon}
L.~Leon, M.~S.~Cavilla, M.~B.~Doran, F.~M.~Warren, J.~Abbott, Scala-tympani
  phantom with cochleostomy and round-window openings for cochlear-implant
  insertion experiments, Journal of Medical Devices 8 (2014) 041010.
\newblock \href {http://dx.doi.org/10.1115/1.4027617}
  {\path{doi:10.1115/1.4027617}}.

\bibitem{DHANASINGH201793}
A.~Dhanasingh, C.~Jolly,
  \href{http://www.sciencedirect.com/science/article/pii/S0378595517302940}{An
  overview of cochlear implant electrode array designs}, Hearing Research 356
  (2017) 93 -- 103.
\newblock \href
  {http://dx.doi.org/https://doi.org/10.1016/j.heares.2017.10.005}
  {\path{doi:https://doi.org/10.1016/j.heares.2017.10.005}}.
\newline\urlprefix\url{http://www.sciencedirect.com/science/article/pii/S0378595517302940}

\bibitem{medel}
MED-EL, Ensuring a hearing future, only with med-el electrodes,
  https://www.medel.com/ca/electrodes/ (Last access March 2020).

\bibitem{majdani}
O.~Majdani, M.~Leinung, T.~Rau, A.~Akbarian, M.~Zimmerling, M.~Lenarz,
  T.~Lenarz, R.~Labadie, Demagnetization of cochlear implants and temperature
  changes in {3.0T MRI} environment, Otolaryngology--head and neck surgery :
  official journal of American Academy of Otolaryngology-Head and Neck Surgery
  139 (2008) 833--9.
\newblock \href {http://dx.doi.org/10.1016/j.otohns.2008.07.026}
  {\path{doi:10.1016/j.otohns.2008.07.026}}.

\bibitem{Kassemi:2005}
M.~Kassemi, D.~Deserranno, J.~G. Oas,
  \href{http://dx.doi.org/10.1016/j.compstruc.2004.08.001}{Fluid-structural
  interactions in the inner ear}, Comput. Struct. 83~(2-3) (2005) 181--189.
\newblock \href {http://dx.doi.org/10.1016/j.compstruc.2004.08.001}
  {\path{doi:10.1016/j.compstruc.2004.08.001}}.
\newline\urlprefix\url{http://dx.doi.org/10.1016/j.compstruc.2004.08.001}

\bibitem{Baertschi1975}
A.~J. Baertschi, R.~N. Johnson, G.~R. Hanna,
  \href{https://doi.org/10.1007/BF00342638}{A theoretical and experimental
  determination of vestibular dynamics in caloric stimulation}, Biological
  Cybernetics 20~(3) (1975) 175--186.
\newblock \href {http://dx.doi.org/10.1007/BF00342638}
  {\path{doi:10.1007/BF00342638}}.
\newline\urlprefix\url{https://doi.org/10.1007/BF00342638}

\bibitem{Cawthorne}
T.~Cawthorne, W.~A. Cobb,
  \href{https://doi.org/10.3109/00016485409127670}{Temperature changes in the
  perilymph space in response to caloric stimulation in man}, Acta
  Oto-Laryngologica 44~(5-6) (1954) 580--588.
\newblock \href
  {http://arxiv.org/abs/https://doi.org/10.3109/00016485409127670}
  {\path{arXiv:https://doi.org/10.3109/00016485409127670}}, \href
  {http://dx.doi.org/10.3109/00016485409127670}
  {\path{doi:10.3109/00016485409127670}}.
\newline\urlprefix\url{https://doi.org/10.3109/00016485409127670}

\bibitem{morshed}
M.~Szymański, K.~Morshed, R.~Mills, Experimental study on heat transmission to
  the vestibule during co2 laser use in revision stapes surgery, The Journal of
  Laryngology and Otology 121~(1) (2007) 5–8.
\newblock \href {http://dx.doi.org/10.1017/S0022215106002672}
  {\path{doi:10.1017/S0022215106002672}}.

\bibitem{ricci_mazzoni_1985}
T.~Ricci, M.~Mazzoni, Experimental investigation of temperature gradients in
  the inner ear following argon laser exposure, The Journal of Laryngology and
  Otology 99~(4) (1985) 359–362.
\newblock \href {http://dx.doi.org/10.1017/S0022215100096833}
  {\path{doi:10.1017/S0022215100096833}}.

\bibitem{Kodali}
S.~Kodali, S.~A. Harvey, T.~E. Prieto,
  \href{https://onlinelibrary.wiley.com/doi/abs/10.1097/00005537-199711000-00005}{Thermal
  effects of laser stapedectomy in an animal model: Co2 versus ktp}, The
  Laryngoscope 107~(11) (1997) 1445--1450.
\newblock \href
  {http://arxiv.org/abs/https://onlinelibrary.wiley.com/doi/pdf/10.1097/00005537-199711000-00005}
  {\path{arXiv:https://onlinelibrary.wiley.com/doi/pdf/10.1097/00005537-199711000-00005}},
  \href {http://dx.doi.org/10.1097/00005537-199711000-00005}
  {\path{doi:10.1097/00005537-199711000-00005}}.
\newline\urlprefix\url{https://onlinelibrary.wiley.com/doi/abs/10.1097/00005537-199711000-00005}

\bibitem{McIntosh}
R.~L. McIntosh, S.~Iskra, R.~J. McKenzie, J.~Chambers, B.~Metzenthen,
  V.~Anderson,
  \href{https://onlinelibrary.wiley.com/doi/abs/10.1002/bem.20364}{Assessment
  of sar and thermal changes near a cochlear implant system for mobile phone
  type exposures}, Bioelectromagnetics 29~(1) (2008) 71--80.
\newblock \href
  {http://arxiv.org/abs/https://onlinelibrary.wiley.com/doi/pdf/10.1002/bem.20364}
  {\path{arXiv:https://onlinelibrary.wiley.com/doi/pdf/10.1002/bem.20364}},
  \href {http://dx.doi.org/10.1002/bem.20364} {\path{doi:10.1002/bem.20364}}.
\newline\urlprefix\url{https://onlinelibrary.wiley.com/doi/abs/10.1002/bem.20364}

\bibitem{Bernardi}
P.~{Bernardi}, M.~{Cavagnaro}, S.~{Pisa}, E.~{Piuzzi}, Specific absorption rate
  and temperature increases in the head of a cellular-phone user, IEEE
  Transactions on Microwave Theory and Techniques 48~(7) (2000) 1118--1126.
\newblock \href {http://dx.doi.org/10.1109/22.848494}
  {\path{doi:10.1109/22.848494}}.

\bibitem{McIntosh2005}
R.~L. McIntosh, V.~Anderson, R.~J. McKenzie,
  \href{https://onlinelibrary.wiley.com/doi/abs/10.1002/bem.20112}{A numerical
  evaluation of sar distribution and temperature changes around a metallic
  plate in the head of a rf exposed worker}, Bioelectromagnetics 26~(5) (2005)
  377--388.
\newblock \href
  {http://arxiv.org/abs/https://onlinelibrary.wiley.com/doi/pdf/10.1002/bem.20112}
  {\path{arXiv:https://onlinelibrary.wiley.com/doi/pdf/10.1002/bem.20112}},
  \href {http://dx.doi.org/10.1002/bem.20112} {\path{doi:10.1002/bem.20112}}.
\newline\urlprefix\url{https://onlinelibrary.wiley.com/doi/abs/10.1002/bem.20112}

\bibitem{Wanger}
F.~Wagner, W.~Wimmer, L.~Leidolt, M.~Vischer, S.~Weder, R.~Wiest,
  G.~Mantokoudis, M.~D. Caversaccio,
  \href{https://doi.org/10.1371/journal.pone.0132483}{Significant artifact
  reduction at 1.5t and 3t mri by the use of a cochlear implant with removable
  magnet: An experimental human cadaver study}, PLOS ONE 10~(7) (2015) 1--17.
\newblock \href {http://dx.doi.org/10.1371/journal.pone.0132483}
  {\path{doi:10.1371/journal.pone.0132483}}.
\newline\urlprefix\url{https://doi.org/10.1371/journal.pone.0132483}

\bibitem{LOEFFLER2007583}
K.~A. Loeffler, T.~A. Johnson, R.~A. Burne, P.~J. Antonelli,
  \href{http://www.sciencedirect.com/science/article/pii/S019459980603436X}{Biofilm
  formation in an in vitro model of cochlear implants with removable magnets},
  Otolaryngology - Head and Neck Surgery 136~(4) (2007) 583 -- 588.
\newblock \href
  {http://dx.doi.org/https://doi.org/10.1016/j.otohns.2006.11.005}
  {\path{doi:https://doi.org/10.1016/j.otohns.2006.11.005}}.
\newline\urlprefix\url{http://www.sciencedirect.com/science/article/pii/S019459980603436X}

\bibitem{YUN2005275}
J.~M. Yun, M.~W. Colburn, P.~J. Antonelli,
  \href{http://www.sciencedirect.com/science/article/pii/S0194599805003864}{Cochlear
  implant magnet displacement with minor head trauma}, Otolaryngology - Head
  and Neck Surgery 133~(2) (2005) 275 -- 277.
\newblock \href
  {http://dx.doi.org/https://doi.org/10.1016/j.otohns.2005.02.018}
  {\path{doi:https://doi.org/10.1016/j.otohns.2005.02.018}}.
\newline\urlprefix\url{http://www.sciencedirect.com/science/article/pii/S0194599805003864}

\bibitem{THOMPSON201546}
A.~C. Thompson, J.~B. Fallon, A.~K. Wise, S.~A. Wade, R.~K. Shepherd, P.~R.
  Stoddart,
  \href{http://www.sciencedirect.com/science/article/pii/S0378595515000635}{Infrared
  neural stimulation fails to evoke neural activity in the deaf guinea pig
  cochlea}, Hearing Research 324 (2015) 46 -- 53.
\newblock \href
  {http://dx.doi.org/https://doi.org/10.1016/j.heares.2015.03.005}
  {\path{doi:https://doi.org/10.1016/j.heares.2015.03.005}}.
\newline\urlprefix\url{http://www.sciencedirect.com/science/article/pii/S0378595515000635}

\bibitem{Shiparo}
M.~Shapiro, K.~Homma, S.~Villarreal, C.-P. Richter, F.~Bezanilla, Infrared
  light excites cells by changing their electrical capacitance, Nature
  communications 3 (2012) 736.
\newblock \href {http://dx.doi.org/10.1038/ncomms1742}
  {\path{doi:10.1038/ncomms1742}}.

\bibitem{Izzo2007OpticalPV}
A.~D. Izzo, J.~T. Walsh, E.~D. Jansen, M.~Bendett, J.~Webb, H.~Ralph, C.-P.
  Richter, Optical parameter variability in laser nerve stimulation: A study of
  pulse duration, repetition rate, and wavelength, IEEE Transactions on
  Biomedical Engineering 54 (2007) 1108--1114.

\bibitem{Thompson2013InfraredNS}
A.~C. Thompson, S.~A. Wade, N.~C. Pawsey, P.~R. Stoddart, Infrared neural
  stimulation: Influence of stimulation site spacing and repetition rates on
  heating, IEEE Transactions on Biomedical Engineering 60 (2013) 3534--3541.

\bibitem{rajguru4-INS}
X.~Tan, S.~Rajguru, H.~Young, N.~Xia, S.~R.~Stock, X.~Xiao, C.~P. Richter,
  Radiant energy required for infrared neural stimulation, Scientific Reports
  5~(13273 (2015)).
\newblock \href {http://dx.doi.org/https://doi.org/10.1038/srep13273}
  {\path{doi:https://doi.org/10.1038/srep13273}}.

\bibitem{Rajguru}
I.~Tamames, C.~King, C.~Y. Huang, M.~E. Telischi, F. F.and~Hoffer, S.~M.
  Rajguru, Theoretical evaluation and experimental validation of localized
  therapeutic hypothermia application to preserve residual hearing after
  cochlear implantation., Ear and hearing 39~(4) (2018) 712--719.
\newblock \href {http://dx.doi.org/doi: 10.1097/AUD.0000000000000529}
  {\path{doi:doi: 10.1097/AUD.0000000000000529}}.

\bibitem{rajguru2-terapeutic}
E.~Perez, A.~Viziano, Z.~Al-Zaghal, F.~F. Telischi, R.~Sangaletti, W.~Jiang,
  W.~D. Dietrich, C.~King, M.~E. Hoffer, S.~M. Rajguru, Anatomical correlates
  and surgical considerations for localized therapeutic hypothermia application
  in cochlear implantation surgery., Otology $\&$ neurotology : official
  publication of the American Otological Society, American Neurotology Society
  [and] European Academy of Otology and Neurotology 40~(9) (2019) 1167--1177.
\newblock \href
  {http://dx.doi.org/https://doi.org/10.1097/MAO.0000000000002373}
  {\path{doi:https://doi.org/10.1097/MAO.0000000000002373}}.

\bibitem{rajguru3-therapeutic}
I.~Tamames, C.~King, E.~Bas, W.~D. Dietrich, F.~Telischi, S.~M. Rajguru,
  \href{http://www.sciencedirect.com/science/article/pii/S0378595516300880}{A
  cool approach to reducing electrode-induced trauma: Localized therapeutic
  hypothermia conserves residual hearing in cochlear implantation}, Hearing
  Research 339 (2016) 32 -- 39.
\newblock \href
  {http://dx.doi.org/https://doi.org/10.1016/j.heares.2016.05.015}
  {\path{doi:https://doi.org/10.1016/j.heares.2016.05.015}}.
\newline\urlprefix\url{http://www.sciencedirect.com/science/article/pii/S0378595516300880}

\bibitem{Perilymph}
R.~Murray, W.~Potts,
  \href{http://www.sciencedirect.com/science/article/pii/0010406X61900731}{The
  composition of the endolymph, perilymph}, Comparative Biochemistry and
  Physiology 2~(1) (1961) 65 -- 75.
\newblock \href
  {http://dx.doi.org/https://doi.org/10.1016/0010-406X(61)90073-1}
  {\path{doi:https://doi.org/10.1016/0010-406X(61)90073-1}}.
\newline\urlprefix\url{http://www.sciencedirect.com/science/article/pii/0010406X61900731}

\bibitem{petruska}
A.~J. Petruska, J.~J. Abbott, Omnimagnet: An omnidirectional electromagnet for
  controlled dipole-field generation, IEEE TRANSACTIONS ON MAGNETICS 50~(7)
  (2014) 1--10.
\newblock \href {http://dx.doi.org/10.1109/TMAG.2014.2303784}
  {\path{doi:10.1109/TMAG.2014.2303784}}.

\bibitem{bone}
H.~Bowman, \href{https://doi.org/10.1080/16070658.1981.11689231}{Heat transfer
  and thermal dosimetry}, Journal of Microwave Power 16~(2) (1981) 121--133.
\newblock \href
  {http://arxiv.org/abs/https://doi.org/10.1080/16070658.1981.11689231}
  {\path{arXiv:https://doi.org/10.1080/16070658.1981.11689231}}, \href
  {http://dx.doi.org/10.1080/16070658.1981.11689231}
  {\path{doi:10.1080/16070658.1981.11689231}}.
\newline\urlprefix\url{https://doi.org/10.1080/16070658.1981.11689231}

\bibitem{Bergheau}
J.~Bargheau, R.~Fortunier, Finite Element Simulation of Heat Transfer, Wiley,
  2008.
\newblock \href {http://dx.doi.org/10.1002/9780470611418}
  {\path{doi:10.1002/9780470611418}}.

\bibitem{reddy}
J.~Reddy, D.~Gartling,
  \href{https://books.google.com/books?id=h7TMBQAAQBAJ}{The Finite Element
  Method in Heat Transfer and Fluid Dynamics}, Applied and Computational
  Mechanics, CRC Press, 2010.
\newline\urlprefix\url{https://books.google.com/books?id=h7TMBQAAQBAJ}

\bibitem{comsol}
\href{https://cdn.comsol.com/doc/5.5/ApplicationProgrammingGuide.pdf}{COMSOL
  Multiphysics, Application Programming Guide}, COMSOL Multiphysics$^{®}$ v.
  5.4, 2019.
\newline\urlprefix\url{https://cdn.comsol.com/doc/5.5/ApplicationProgrammingGuide.pdf}

\bibitem{ozisik}
M.~Ozisik, \href{https://books.google.com/books?id=WyZRAAAAMAAJ}{Heat
  Conduction}, Wiley, 1993.
\newline\urlprefix\url{https://books.google.com/books?id=WyZRAAAAMAAJ}

\bibitem{Jain}
P.~Jain, S.~Singh, R.~Uddin, Analytical solution to transient asymmetric heat
  conduction in a multilayer annulus, Journal of Heat Transfer 131~(1) (2009)
  1--7.
\newblock \href {http://dx.doi.org/10.1115/1.2977553}
  {\path{doi:10.1115/1.2977553}}.

\bibitem{cho}
C.~H. Cho, K.~S. Chang, K.~H. Park, Numerical simulation of natural convection
  in concentric and eccentric horizontal cylindrical annuli, ASME Transactions
  Journal of Heat Transfer 104 (1982) 624--630.
\newblock \href {http://dx.doi.org/10.1115/1.3245177}
  {\path{doi:10.1115/1.3245177}}.

\bibitem{kuehn}
T.~Kuehn, R.~Goldstein, Correlating equations for natural convection heat
  transfer between horizontal circular cylinders, International Journal of Heat
  and Mass Transfer 19~(10) (1976) 1127--1134.
\newblock \href {http://dx.doi.org/10.1016/0017-9310(76)90145-9}
  {\path{doi:10.1016/0017-9310(76)90145-9}}.

\bibitem{reviewer-paper1}
A.~Fasquelle, J.~Pelle, S.~Harmand, I.~V. Shevchuk, Numerical study of
  convective heat transfer enhancement in a pipe rotating around a parallel
  axis, Journal of Heat Transfer 19~(136(5)) (2014) 051901.
\newblock \href {http://dx.doi.org/https://doi.org/10.1115/1.4025642}
  {\path{doi:https://doi.org/10.1115/1.4025642}}.

\bibitem{reviewer-paper2}
W.~Seddique, L.~El-Gabry, I.~V.Shevchuk, N.~B. Hushmandi, T.~H. Fransson, Flow
  structure, heat transfer and pressure drop in varying aspect ratio two-pass
  rectangular smooth channels, Heat Mass Transfer 48 (2011) 735--748.
\newblock \href {http://dx.doi.org/https://doi.org/10.1007/s00231-011-0926-1}
  {\path{doi:https://doi.org/10.1007/s00231-011-0926-1}}.

\bibitem{reviewer-paper3}
W.~Siddique, I.~V. Shevchuk, L.~El-Gabry, N.~B. Hushmandi, T.~H. Fransson, On
  flow structure, heat transfer and pressure drop in varying aspect ratio
  two-pass rectangular channel with ribs at {45$^\circ$}, Heat Mass Transfer
  19~(49) (2013) 679--694.
\newblock \href {http://dx.doi.org/10.1016/0017-9310(76)90145-9}
  {\path{doi:10.1016/0017-9310(76)90145-9}}.

\bibitem{thermaldose1}
P.~S. Yarmolenko, E.~J. Moon, C.~Landon, A.~Manzoor, D.~W. Hochman, B.~L.
  Viglianti, M.~W. Dewhirst,
  \href{https://doi.org/10.3109/02656736.2010.534527}{Thresholds for thermal
  damage to normal tissues: An update}, International Journal of Hyperthermia
  27~(4) (2011) 320--343, pMID: 21591897.
\newblock \href
  {http://arxiv.org/abs/https://doi.org/10.3109/02656736.2010.534527}
  {\path{arXiv:https://doi.org/10.3109/02656736.2010.534527}}, \href
  {http://dx.doi.org/10.3109/02656736.2010.534527}
  {\path{doi:10.3109/02656736.2010.534527}}.
\newline\urlprefix\url{https://doi.org/10.3109/02656736.2010.534527}

\bibitem{Yoshida10116}
N.~Yoshida, A.~Kristiansen, M.~C. Liberman,
  \href{http://www.jneurosci.org/content/19/22/10116}{Heat stress and
  protection from permanent acoustic injury in mice}, Journal of Neuroscience
  19~(22) (1999) 10116--10124.
\newblock \href
  {http://arxiv.org/abs/http://www.jneurosci.org/content/19/22/10116.full.pdf}
  {\path{arXiv:http://www.jneurosci.org/content/19/22/10116.full.pdf}}, \href
  {http://dx.doi.org/10.1523/JNEUROSCI.19-22-10116.1999}
  {\path{doi:10.1523/JNEUROSCI.19-22-10116.1999}}.
\newline\urlprefix\url{http://www.jneurosci.org/content/19/22/10116}

\bibitem{Rhoon}
G.~C. Van~Rhoon, S.~Theodoros, P.~S. Yarmolenko, M.~W. Dewhirst, E.~Neufeld,
  N.~Kuster, Cem$_{43}$ thermal dose thresholds: a potential guide for magnetic
  resonance radiofrequency exposure levels?, European Radiology 23~(8) (2013)
  2215--2227.
\newblock \href {http://dx.doi.org/https://doi.org/10.1007/s00330-013-2825-y}
  {\path{doi:https://doi.org/10.1007/s00330-013-2825-y}}.

\bibitem{incropera}
T.~Bergman, A.~Lavine, F.~Incropera,
  \href{https://books.google.com/books?id=5cgbAAAAQBAJ}{Fundamentals of Heat
  and Mass Transfer, 7th Edition}, John Wiley $\&$ Sons, Incorporated, 2011.
\newline\urlprefix\url{https://books.google.com/books?id=5cgbAAAAQBAJ}

\bibitem{RAITHBY}
G.~Raithby, K.~Hollands,
  \href{http://www.sciencedirect.com/science/article/pii/S0065271708700765}{A
  general method of obtaining approximate solutions to laminar and turbulent
  free convection problems}, Advances in Heat Transfer 11 (1975) 265 -- 315.
\newblock \href
  {http://dx.doi.org/https://doi.org/10.1016/S0065-2717(08)70076-5}
  {\path{doi:https://doi.org/10.1016/S0065-2717(08)70076-5}}.
\newline\urlprefix\url{http://www.sciencedirect.com/science/article/pii/S0065271708700765}

\bibitem{lubricant}
G.~Kontorinis, G.~Paasche, T.~lenarz, T.~St\"{o}ver, The effect of different
  lubricants on cochlear implant electrode insertion forces, Otology $\&$
  Neurotology 32~(7) (2011) 1050--1056.
\newblock \href {http://dx.doi.org/10.1097/MAO.0b013e31821b3c88}
  {\path{doi:10.1097/MAO.0b013e31821b3c88}}.

\bibitem{glycerol}
A.~Singh, R.~Walvekar, M.~Khalid, W.~Yin~Wong, T.~Gupta, Thermophysical
  properties of glycerol and polyethylene glycol (peg 600) based des, Journal
  of Molecular Liquids 252 (2018) 439--444.
\newblock \href
  {http://dx.doi.org/https://doi.org/10.1016/j.molliq.2017.10.030}
  {\path{doi:https://doi.org/10.1016/j.molliq.2017.10.030}}.

\bibitem{glycerolcp}
D.~Jahn, F.~Akinkunmi, N.~Giovambattista, Effects of temperature on the
  properties of glycerol: A computer simulation study of five different force
  fields, The Journal of Physical Chemistry B 118~(38) (2014) 11284--11294.
\newblock \href {http://dx.doi.org/10.1021/jp5059098}
  {\path{doi:10.1021/jp5059098}}.

\bibitem{soap}
D.~Glover, T.~a. Lomer, Some thermodynamic properties of potassium soaps,
  Molecular Crystals and Liquid Crystals 53 (1979) 181--188.
\newblock \href {http://dx.doi.org/https://doi.org/10.1080/00268947908083994}
  {\path{doi:https://doi.org/10.1080/00268947908083994}}.

\bibitem{soap1}
M.~Zhou, G.~Xia, J.~Li, L.~Chai, L.~Zhou, Analysis of factors influencing
  thermal conductivity and viscosity in different kinds of surfactant
  solutions, Experimental Thermal and Fluid Science 36 (2012) 22--29.
\newblock \href {http://dx.doi.org/10.1016/j.expthermflusci.2011.07.014}
  {\path{doi:10.1016/j.expthermflusci.2011.07.014}}.

\end{thebibliography}

\end{document}